\documentclass[10pt,journal,cspaper,compsoc]{IEEEtran}

\usepackage{amssymb}
\usepackage{amsmath}
\usepackage{multirow}
\usepackage{algorithm}
\usepackage{algorithmic}
\usepackage{graphicx}

\newcommand{\partitle}[1]{\medskip \noindent \textbf{#1.}}

\begin{document}
\graphicspath{{figures/}}
\title{DPCube: Differentially Private Histogram Release through Multidimensional Partitioning}

\author{Yonghui Xiao,
        Li Xiong,
        Liyue Fan
        and Slawomir Goryczka
\IEEEcompsocitemizethanks{
\IEEEcompsocthanksitem{A preliminary version of the manuscript appeared in \cite{SDMpaper} and a demonstration description will appear in \cite{xiao12icde}}
\IEEEcompsocthanksitem Y. Xiao, L. Xiong, L. Fan. and S. Goryczka are with the Department of Mathematics \& Computer Science, Emory University, Atlanta, GA, 30322.\protect\\
E-mail: \{yonghui.xiao, lxiong, lfan3, sgorycz\}@emory.edu
\IEEEcompsocthanksitem This research was supported in part by NSF grant CNS-1117763, a Cisco Research Award, and an Emory URC grant.
}
} 

\IEEEcompsoctitleabstractindextext{%
\begin{abstract}
Differential privacy is a strong notion for protecting individual privacy in privacy preserving data analysis or publishing.
In this paper, we study the problem of differentially private histogram release for random workloads.  We study two multidimensional partitioning strategies including: 1) a baseline cell-based partitioning strategy for releasing an equi-width cell histogram, and 2) an innovative 2-phase kd-tree based partitioning strategy for releasing a v-optimal histogram.  We formally analyze the utility of the released histograms and quantify the errors for answering linear queries such as counting queries.  We formally characterize the property of the input data that will guarantee the optimality of the algorithm.  Finally, we implement and experimentally evaluate several applications using the released histograms, including counting queries, classification, and blocking for record linkage and show the benefit of our approach.

\end{abstract}

}
\maketitle
\IEEEdisplaynotcompsoctitleabstractindextext
\IEEEpeerreviewmaketitle

\newtheorem{theorem1}{Theorem}[section]
\newtheorem{Lemma}{Lemma}[section]
\newtheorem{definition1}{Definition}[section]
\newtheorem{corollary1}{Corollary}[section]
\newtheorem{observation1}{Observation}[section]

\section{Introduction}

As information technology enables the collection, storage, and usage of massive amounts of information
about individuals and organizations, privacy becomes an increasingly important issue.  Governments and organizations recognize the critical value in sharing such information while preserving the privacy of individuals.  Privacy preserving data analysis and data publishing \cite{DBLP:conf/tamc/Dwork08,FWCY10csur,cacm} has received considerable attention in recent years.
There are two models for privacy protection \cite{DBLP:conf/tamc/Dwork08}: the interactive model and the non-interactive model.
In the interactive model, a trusted {\em curator} (e.g. hospital)
collects data from {\em record owners} (e.g. patients) and provides an access mechanism for {\em data
users} (e.g. public health researchers) for querying or analysis purposes.  The result returned from
the access mechanism is perturbed by the mechanism to protect privacy.  In the non-interactive model, the curator publishes a
``sanitized'' version of the data,
simultaneously providing utility for data users and privacy protection for the individuals represented in the data.

Differential privacy
\cite{Dwork-calibrating,Dwork-dp,DBLP:conf/tamc/Dwork08,cacm,Kifer-no-free-lunch}
is widely accepted as one of the strongest known privacy guarantees
with the advantage that it makes few assumptions on the attacker's
background knowledge. It requires the outcome of computations to be
formally indistinguishable when run with or without any particular
record in the dataset, as if it makes little difference whether an
individual is being opted in or out of the database. Many meaningful
results have been obtained for the interactive model with
differential privacy
\cite{Dwork-calibrating,Dwork-dp,DBLP:conf/tamc/Dwork08,cacm}.
Non-interactive data release with differential privacy has been
recently studied with hardness results obtained and it remains an
open problem to find efficient algorithms for many domains
\cite{Blum-learningapproach,Dwork-complexity}.

\vspace{-0.3cm}
\begin{figure}[h!]
\vspace{-0.5cm}
\includegraphics[width=9.5cm]{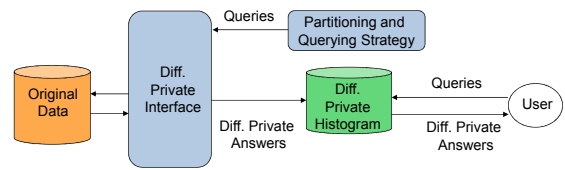}
\vspace{-4.5cm}
\caption{Differentially private histogram release}
\label{Figure-release}
\end{figure}

In this paper, we study the problem of differentially private histogram release based on a differential privacy interface, as shown in Figure \ref{Figure-release}.  A {\em histogram} is a disjoint partitioning of the database points with the number of points which fall into each partition.  A differential privacy interface, such as the Privacy INtegrated Queries platform (PINQ) \cite{McSherry-PINQ}, provides a differentially private access to the raw database.  An algorithm implementing the partitioning strategy submits a sequence of queries to the interface and generates a differentially private histogram of the raw database.  The histogram can then serve as a sanitized synopsis of the raw database and, together with an optional synthesized dataset based on the histogram, can be used to support count queries and other types of OLAP queries and learning tasks.

An immediate question one might wonder is what is the advantage of the non-interactive release compared to using the interactive mechanism to answer the queries directly.  A common mechanism providing differentially private answers is to add carefully calibrated noise to each query determined by the privacy parameter and the sensitivity of the query.  The composability of differential privacy \cite{McSherry-PINQ} ensures privacy guarantees for a sequence of differentially-private computations with additive privacy depletions in the worst case.  Given an overall privacy requirement or budget, expressed as a privacy parameter, it can be allocated to subroutines or each query in the query sequence to ensure the overall privacy.   When the number of queries grow, each query gets a lower privacy budget which requires a larger noise to be added.  When there are multiple users, they have to share a common privacy budget which degrades the utility rapidly.  The non-interactive approach essentially exploits the data distribution and the query workload and uses a carefully designed algorithm or query strategy such that the overall noise is minimized for a particular class of queries.  As a result, the partitioning strategy and the algorithm implementing the strategy for generating the query sequence to the interface are crucial to the utility of the resulting histogram or synthetic dataset.  

\smallskip
\noindent
{\bf Contributions}.
We study differentially private histogram release for random query workload in this paper and propose partitioning and estimation algorithms with formal utility analysis and experimental evaluations. We summarize our contributions below.

\begin{itemize}
\item
We study two multidimensional partitioning strategies for differentially private histogram release: 1) a baseline cell-based partitioning strategy for releasing an equi-width cell histogram, and 2) an innovative 2-phase kd-tree (k-dimensional tree) based space partitioning strategy for releasing a v-optimal histogram.  There are several innovative features in our 2-phase strategy. First, we incorporate a uniformity measure in the partitioning process which seeks to produce partitions that are close to uniform so that approximation errors within partitions are minimized, essentially resulting in a differentially private v-optimal histogram.  Second, we implement the strategy using a two-phase algorithm that generates the kd-tree partitions based on the cell histogram so that the access to the differentially private interface is minimized.
\item
We formally analyze the utility of the released histograms and quantify the errors for answering linear distribute queries such as counting queries.  We show that the cell histogram provides bounded query error for any input data.  We also show that the v-optimal histogram combined with a simple query estimation scheme achieves bounded query error and superior utility than existing approaches for ``smoothly" distributed data.  We formally characterize the ``smoothness" property of the input data that guarantees the optimality of the algorithm.
\item
We implement and experimentally evaluate several applications using the released histograms, including counting queries, classification, and blocking for record linkage. We compare our approach with other existing privacy-preserving algorithms and show the benefit of our approach.
\end{itemize}

\section{Related works}
Privacy preserving data analysis and publishing has received considerable attention in recent years.
We refer readers to \cite{DBLP:conf/tamc/Dwork08,FWCY10csur,cacm} for several up-to-date surveys.
We briefly review here the most relevant work to our paper and discuss how our work differs from existing work.

There has been a series of studies on interactive privacy preserving data analysis based on
the notion of differential privacy \cite{Dwork-calibrating,Dwork-dp,DBLP:conf/tamc/Dwork08,cacm}.
A primary approach proposed for achieving differential privacy is to add Laplace noise \cite{Dwork-calibrating,DBLP:conf/tamc/Dwork08,Dwork-dp} to the original results.
McSherry and Talwar \cite{McSherry-mechanism} give an alternative method to implement differential privacy based on the probability of a returned result, called the exponential mechanism.
Roth and Roughgarden \cite{DBLP:conf/stoc/RothR10} proposes a median mechanism which improves upon the Laplace mechanism.
McSherry implemented the interactive data access mechanism into PINQ \cite{McSherry-PINQ}, a platform providing a programming interface through a SQL-like language.

A few works started addressing non-interactive data release that achieves differential privacy.  Blum et al. \cite{Blum-learningapproach} proved the possibility of
non-interactive data release satisfying differential privacy for
queries with polynomial VC-dimension, such as predicate queries. It
also proposed an inefficient algorithm based on the exponential
mechanism.  The result largely remains theoretical and the general
algorithm is inefficient for the complexity and required data size.  \cite{Dwork-complexity} further proposed more efficient algorithms
with hardness results obtained and it remains a key open problem to
find efficient algorithms for non-interactive data release with
differential privacy for many domains.
\cite{HR10} pointed out that a natural approach to side-stepping the hardness is relaxing the utility requirement, and not requiring accuracy for {\em every} input database.

Several recent work studied differentially private mechanisms for
particular kinds of data such as search logs
\cite{citeulike:4387902,DBLP:journals/corr/abs-0904-0682} or set-valued data \cite{set-valued-vldb2011}.  Others proposed algorithms for specific applications or optimization goals such as recommender systems \cite{1557090}, record linkage \cite{inan10private}, data mining \cite{Mohammed:2011:DPD:2020408.2020487}, or differentially private data cubes with minimized overall cuboid error \cite{Ding:2011:DPD:1989323.1989347}. It is important to note that
\cite{inan10private} uses several tree strategies including kd-tree
in its partitioning step and our results show that our 2-phase uniformity-driven kd-tree strategy achieves better utility for random count queries.

A few works considered releasing data for predictive count queries and are closely related to ours.  X. Xiao et al. \cite{DBLP:conf/icde/XiaoWG10} developed an algorithm using wavelet transforms.  \cite{boost-accuracy} generates differentially private histograms for single dimensional range queries through a hierarchical partitioning approach and a consistency check technique.
\cite{Optimizing-PODS} proposes a query matrix mechanism that generates an optimal query strategy based on the query workload of linear count queries and further mapped the work in \cite{DBLP:conf/icde/XiaoWG10} and \cite{boost-accuracy} as special query strategies that can be represented by a query matrix.  It is worth noting that the cell-based
partitioning in our approach is essentially the identity query
matrix referred in \cite{Optimizing-PODS}.  While we will leverage the query matrix framework to formally analyze our approach, it is important to note that the above mentioned query strategies are data-oblivious in that they are determined by the query workload, a static wavelet matrix, or hierarchical matrix without taking into consideration the underlying data.  On the other hand, our 2-phase kd-tree based partitioning is designed to explicitly exploit the smoothness of the underlying data indirectly observed by the differentially private interface and the final query matrix corresponding to the released histogram is dependent on the approximate data distribution.  We are also aware of a forthcoming work \cite{xu12icde} and will compare with it as the proceedings becomes available.

In summary, our work complements and advances the above works in that we focus on differentially private histogram release for random query workload using a multidimensional partitioning approach that is ``data-aware".
Sharing the insights from \cite{HR10,set-valued-vldb2011}, our primary viewpoint is that it is possible and desirable in practice to design adaptive or data-dependent heuristic mechanisms for differentially private data release for useful families or subclasses of databases and applications.
Our approach provides formal utility guarantees for a class of queries and also supports a variety of applications including general OLAP, classification and record linkage.

\section{Preliminaries and Definitions}


In this section, we formally introduce the definitions of differential privacy, the data model and the queries we consider, as well as a formal utility notion called $(\epsilon,\delta)$-usefulness.  Matrices and vectors are indicated with bold letters (e.g $\textbf{H}$, $\textbf{x}$) and their elements are indicated as $H_{ij}$ or $x_i$.  While we will introduce mathematical notations in this section and subsequent sections, Table \ref{tbl-symbols} lists the frequently-used symbols for references.

\begin{table}
\label{tbl-symbols}
\caption{Frequently used symbols}
\begin{tabular}{|c|c|}
\hline
Symbol&Description\\
\hline
$n$ & number of records in the dataset\\
\hline
$m$ & number of cells in the data cube\\
\hline
$x_i$ & original count of cell $i (1 \le i \le m)$\\
\hline
$y_i$ & released count of cell $i$ in cell histogram \\
\hline
$y_p$ & released count of partition $p$ in subcube histogram \\
\hline
$n_p$ & size of partition $p$\\
\hline
$s$ & size of query range\\
\hline
$\alpha$, $\alpha_1$, $\alpha_2$ & differential privacy parameters \\
\hline
$\gamma$ & smoothness parameter\\
\hline
\end{tabular}
\end{table}

\subsection{Differential Privacy}
\begin{definition1}[$\alpha$-Differential privacy \cite{Dwork-dp}]In the interactive model, an access mechanism $\mathcal{A}$ satisfies $\alpha$-differential privacy
if for any neighboring databases\footnote{We use the definition of
unbounded neighboring databases \cite{Kifer-no-free-lunch} consistent with \cite{McSherry-PINQ}
which treats the databases as multisets of records and requires
their symmetric difference to be 1.} $D_1$ and $D_2$, for any query
function $Q$, $r\subseteq Range(Q)$, $\mathcal{A}_Q(D)$ is the
mechanism to return an answer to query $Q(D)$,
\begin{displaymath}
Pr[\mathcal{A}_Q(D_1)=r]\leq
e^{\alpha}Pr[\mathcal{A}_Q(D_2)=r]
\end{displaymath}
In the non-interactive model, a data release mechanism $\mathcal{A}$
satisfies $\alpha$-differential privacy if for all neighboring
database $D_1$ and $D_2$, and released output $\hat{D}$,
\begin{displaymath}
Pr[\mathcal{A}(D_1)=\hat{D}]\leq
e^{\alpha}Pr[\mathcal{A}(D_2)=\hat{D}]
\end{displaymath}
\end{definition1}

\partitle{Laplace Mechanism}
To achieve differential privacy, we use the Laplace mechanism \cite{Dwork-calibrating} that adds random noise of
Laplace distribution to the true answer of a query $Q$,
$\mathcal{A}_Q(D)=Q(D)+\tilde{N}$, where $\tilde{N}$ is the Laplace noise. The magnitude of the noise depends on the privacy level and the query's sensitivity.

\begin{definition1}[Sensitivity]
For arbitrary neighboring databases $D_1$ and $D_2$, the sensitivity of
a query $Q$, denoted by $S_Q$, is the maximum difference between the query results of
$D_1$ and $D_2$,
\begin{equation}
S_Q=max|Q(D_1)-Q(D_2)|
\label{eqn-sensitivity}
\end{equation}
\end{definition1}

To achieve $\alpha$-differential privacy for a given query $Q$ on dataset $D$,
it is sufficient to return $Q(D)+\tilde{N}$ in place of the original result $Q(D)$
where we draw $\tilde{N}$ from $Lap(S_Q/\alpha)$ \cite{Dwork-calibrating}.

\partitle{Composition}
The composability of differential privacy \cite{McSherry-PINQ} ensures privacy guarantees for a sequence of differentially-private computations. For a general series of analysis, the privacy parameter values add up, i.e. the privacy guarantees degrade as we
expose more information.  In a special case that the analyses
operate on disjoint subsets of the data, the ultimate privacy guarantee depends only on the worst of the guarantees of each analysis, not the sum.

\begin{theorem1}[Sequential Composition \cite{McSherry-PINQ}]
\label{lemma-composition}
Let $M_i$ each provide $\alpha_i$-differential privacy.
The sequence of $M_i$ provides $(\sum_i{\alpha_i})$-differential privacy.
\end{theorem1}

\begin{theorem1}[Parallel Composition \cite{McSherry-PINQ}]
\label{lemma-parallel}
If $D_i$ are disjoint subsets of the original database and $M_i$ provides $\alpha$-differential privacy for each $D_i$,
then the sequence of $M_i$ provides $\alpha$-differential privacy.
\end{theorem1}

\partitle{Differential Privacy Interface}
A privacy interface such as PINQ \cite{McSherry-PINQ} can be used to provide a differentially private interface to a database. It provides operators for database aggregate
queries such as count ({\bf NoisyCount}) and sum ({\bf NoisySum}) which uses Laplace noise and
the exponential mechanism to enforce differential privacy.
It also provides a {\bf Partition} operator that can partition the dataset based on the provided set of candidate keys.
The {\bf Partition} operator takes advantage of parallel
composition and thus the privacy costs do not add up.




\subsection{Data and Query Model}
\label{sec-data-cube}

\partitle{Data Model}
Consider a dataset with $N$ nominal or discretized attributes, we use an $N$-dimensional data cube, also called a base cuboid in the data warehousing literature \cite{JWH-dataming,Ding:2011:DPD:1989323.1989347}, to represent the aggregate information of the data set.  The records are the points in the $N$-dimensional data space.  Each cell of a data cube represents an aggregated measure, in our case, the count of the data points corresponding to the multidimensional coordinates of the cell. We denote the number of cells by $m$ and $m = |dom(A_1)| * \dots * |dom(A_N)|$ where $|dom(A_i)|$ is the domain size of attribute $A_i$. We use the term ``partition'' to refer to any sub-cube in the data cube.

\begin{figure}[h!]
\centering
\vspace{-0.8cm}
\includegraphics[width=10cm]{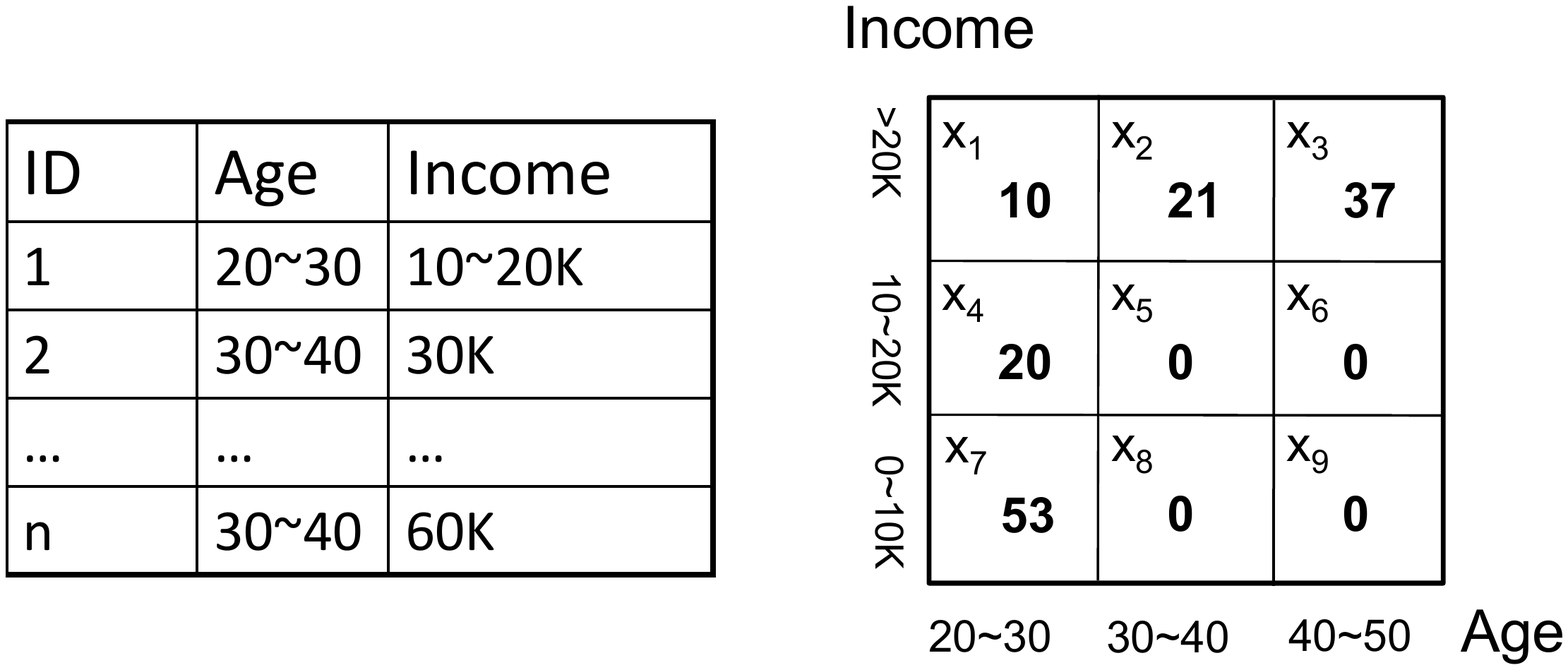}
\vspace{-4cm}
\caption{Running example: original data represented in a relational table (left) and a 2-dimensional count cube (right)}
\label{Fig_exampledata1}
\end{figure}

{\bf Figure \ref{Fig_exampledata1}} shows an example relational dataset with attribute age and income (left) and a two-dimensional count data cube or histogram (right). The domain values of age are $20\scriptsize{\sim}30$, $30\scriptsize{\sim}40$ and $40\scriptsize{\sim}50$; the domain values of income are $0\scriptsize{\sim}10K$, $10K\scriptsize{\sim}20K$ and $>20K$.  Each cell in the data cube represents the population count corresponding to the age and income values.

\partitle{Query Model}
We consider linear counting queries that compute a linear combination of the count values in the data cube based on a query predicate.  We can represent the original data cube, e.g. the counts of all cells, by an $m$-dimensional column vector $\textbf{x}$ shown below.
\begin{small}
\begin{equation*}
\label{eqn_x}
\textbf{x}=
\left[
\begin{array}{ccccccccc}
10&21&37&20&0&0&53&0&0
\end{array}
\right]^T
\end{equation*}
\end{small}
\vspace{-0.5cm}
\begin{definition1}[Linear query \cite{Optimizing-PODS}] A linear query Q can be represented as an $m$-dimensional boolean vector $\textbf{Q} = [q_1 \dots q_n]$ with each $q_i \in \mathbb{R}$.  The answer to a linear query $\textbf{Q}$ on data vector $\textbf{x}$ is the vector product $\textbf{Qx}$ = $q_1x_1 + \dots + q_nx_n$.
\end{definition1}

\begin{figure}[h!]
\centering
\includegraphics[width=9cm]{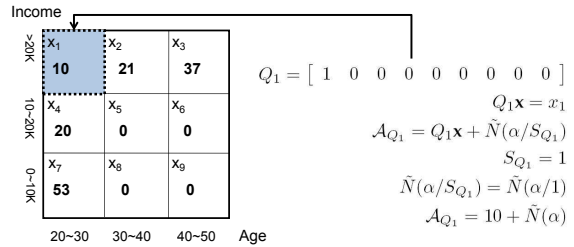}
\vspace{-3.6cm} \caption{Example: a linear counting query}
\label{Fig_model_illu}
\end{figure}

In this paper, we consider counting queries with boolean predicates so that each $q_i$ is a boolean variable with value 0 or 1.  The sensitivity of the counting queries, based on equation (\ref{eqn-sensitivity}), is $S_Q = 1$.  We denote $s$ as the query range size or query size, which is the number of cells contained in the query predicate, and we have $s = |\textbf{Q}|$.  For example, a query $Q_1$ asking the population count with age $= [20,30]$ and income $>30k$, corresponding to $x_1$, is shown as a query vector in Figure \ref{Fig_model_illu}.  
The size of this query is 1.  It also shows the original answer of $Q_1$ and a perturbed answer with Laplace noise that achieves $\alpha$-differential privacy.
We note that the techniques and proofs are generalizable to real number query vectors.

\vspace{-0.4cm}

\begin{definition1} [Query matrix \cite{Optimizing-PODS}] A query matrix is a collection of linear queries, arranged by rows to form an
$p \times m$ matrix.
\end{definition1}
Given a $p \times m$ query matrix \textbf{H}, the query answer for \textbf{H} is a length-$p$ column vector of query results, which can be computed as the matrix product \textbf{Hx}.  For example, an $m \times m$ identity query matrix $\textbf{I}_m$ will result in a length-$m$ column vector consisting of all the cell counts in the original data vector \textbf{x}.

A data release algorithm, consisting of a sequence of designed queries using the differential privacy interface, can be represented as a query matrix.  We will use this query matrix representation in the analysis of our algorithms.

\subsection{Utility Metrics}
We formally analyze the utility of the released data by the notion of $(\epsilon,\delta)$-usefulness \cite{Blum-learningapproach}.
\begin{definition1}[$(\epsilon,\delta$)-usefulness \cite{Blum-learningapproach}]
A database mechanism $\mathcal{A}$ is $(\epsilon,\delta)$-useful for
queries in class C if with probability $1-\delta$, for every $Q\in
C$, and every database D, $\mathcal{A}(D)=\hat{D}$,
$|Q(\hat{D})-Q(D)|\leq\epsilon$.
\end{definition1}

In this paper, we mainly focus on linear counting queries to formally analyze the released histograms.  We will discuss and experimentally show how the released histogram can be used to support other types of OLAP queries such as sum and average and other applications such as classification.


%

\subsection{Laplace Distribution Properties}
We include a general lemma on probability distribution and a theorem on the statistical distribution of the summation of multiple Laplace noises, which we will use when analyzing the utility of our algorithms.

\begin{Lemma}
\label{lemma-PDF-aZ}
If $Z\sim f(z)$, then $aZ\sim \frac{1}{a}f(z/a)$ where $a$ is any constant.
\end{Lemma}

\begin{theorem1}\cite{Bilateral-gamma}
Let $f_{n}(z,\alpha)$ be the PDF of $\sum_{i=1}^{n}\tilde{N_i}(\alpha)$ where $\tilde{N}_i(\alpha)$ are i.i.d. Laplace noise Lap$(1/\alpha)$,
\begin{small}
\begin{multline}
\label{fml-baliteral-gamma}
f_{n}(z,\alpha)=\\\frac{\alpha^{n}}{2^n  \Gamma^2(n)}exp(-{\alpha |z|})\int_0^{\infty}v^{n-1}(|z|+\frac{v}{2\alpha})^{n-1}e^{-v}dv
\end{multline}
\end{small}
\end{theorem1}

\section{Multidimensional Partitioning}

\subsection{Motivation and Overview}
For differentially private histogram release, a multi-dimensional histogram on a set of attributes is constructed by partitioning the data points into mutually disjoint subsets called {\em buckets} or {\em partitions}.  The counts or frequencies in each bucket is then released. Any access to the original database is conducted through the differential privacy interface to guarantee differential privacy.  The histogram can be then used to answer random counting queries and other types of queries.

The partitioning strategy will largely determine the utility of the released histogram to arbitrary counting queries. Each partition introduces a bounded Laplace noise or {\em perturbation error} by the differential privacy interface.
If a query predicate covers multiple partitions, the perturbation error is aggregated.  If a query predicate falls within a partition, the result has to be estimated assuming certain distribution of the data points in the partition.  The dominant approach in histogram literature is making the {\em uniform distribution assumption}, where the frequencies of records in the bucket are assumed to be the same and equal to the average of the actual frequencies \cite{Ioannidis03thehistory}.  This introduces an {\em approximation error}.   


\begin{figure}[h!]
\centering
\vspace{-0.5cm}
\includegraphics[width=9cm]{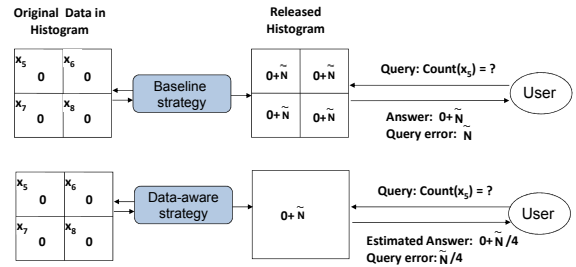}
\vspace{-3.3cm}
\caption{Baseline strategy vs. data-aware strategy}
\label{Figure-DM1}
\end{figure}

\partitle{Example}
We illustrate the errors and the impact of different partitioning strategies through an example shown in Figure \ref{Figure-DM1}. Consider the data in Figure \ref{Fig_exampledata1}. As a baseline strategy, we could release a noisy count for each of the cells.  In a data-aware strategy, as if we know the original data, the 4 cells, $x_5,x_6,x_8,x_9$, can be grouped into one partition and we release a single noisy count for the partition. 
Note that the noise are independently generated for each cell or partition.  Because the sensitivity of the counting query is 1 and the partitioning only requires parallel composition of differential privacy, the magnitude of the noise in the two approaches are the same.
Consider a query, count($x_5$), asking the count of data points in the region $x_5$.  For the baseline strategy, the query error is $\tilde{N}$ which only consists of the perturbation error.
For the data-aware strategy,
the best estimate for the answer based on the uniform distribution assumption is $0+\tilde{N}/4$.  So the query error is $\tilde{N}/4$.  In this case, the approximation error is 0 because the cells in the partition are indeed uniform.  If not, approximation error will be introduced. In addition, the perturbation error is also amortized among the cells. Clearly, the data-aware strategy is desired in this case.

In general, a finer-grained partitioning will introduce smaller approximation errors but larger aggregated perturbation errors.  Finding the right balance to minimize the overall error for a random query workload is a key question. Not surprisingly, finding the optimal multi-dimensional histogram, even without the privacy constraints, is a challenging problem and optimal partitioning even in two dimensions is NP-hard \cite{DBLP:conf/icdt/MuthukrishnanPS99}. Motivated by the above example and guided by the composition theorems, we summarize our two design goals: 1) generate uniform or close to uniform partitions so that the approximation error within the partitions is minimized, essentially generating a v-optimal histogram \cite{Poosala:1996}; 2) carefully and efficiently use the privacy budget to minimize the perturbation error.  In this paper, we first study the most fine-grained cell-based partitioning as a baseline strategy, which results in a equi-width histogram and does not introduce approximation error but only perturbation error.  We then propose a 2-phase kd-tree (k-dimensional tree) based partitioning strategy that results in an v-optimal histogram and seeks to minimize both the perturbation and approximation error.

\subsection{A Baseline Cell Partitioning Strategy}

\begin{figure}[h!]
\centering
\vspace{-0.55cm}
\includegraphics[width=9cm]{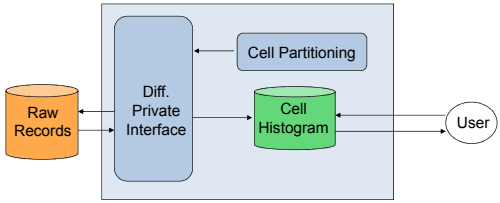}
\vspace{-4.2cm}
\caption{Baseline cell partitioning}
\label{Figure-cell}
\end{figure}

A simple strategy is to partition the data based on the domain and then release a noisy count for each cell which results in a equi-width cell histogram.  Figure \ref{Figure-cell} illustrate this baseline strategy. The implementation is quite simple, taking advantage the {\bf Partition} operator followed by {\bf NoisyCount} on each partition, shown in Algorithm \ref{alg-interval}.

\begin{algorithm}
\caption{Baseline cell partitioning algorithm}
\begin{algorithmic}
\label{alg-interval}
\REQUIRE{
$\alpha$: differential privacy budget
}
\STATE
1. {\bf Partition} the data based on all domains.\\
2. Release {\bf NoisyCount} of each partition using privacy parameter $\alpha$\\
\end{algorithmic}
\end{algorithm}

\partitle{Privacy Guarantee}
We present the theorem below for the cell partitioning algorithm which can be derived directly from the composibility theorems.

\smallskip
\begin{theorem1}%
\label{Theo-dp-Lap}
Algorithm \ref{alg-interval} achieves $\alpha$-differential privacy.
\end{theorem1}
\begin{proof}
Because every cell is a disjoint subset of the original database, according to theorem \ref{lemma-parallel}, it's $\alpha$-differentially private.
\end{proof}

\partitle{Error Quantification}
We present a lemma followed by a theorem that states a formal utility guarantee of cell-based partitioning for linear distributive queries.

\begin{Lemma}
\label{Lemma-SumLap}
If $\tilde{N}_i$ ($ i = 1 \dots m$) is a set of random variables i.i.d from $Lap(b)$ with mean 0, given $0< \epsilon <1$, the following holds:
\begin{equation}
\label{equation-Lemma-SumLap}
Pr[\sum_{i=1}^m |\tilde{N}_i| \leq \epsilon] \geq 1-m\cdot exp(-\frac{\epsilon}{m b})
\end{equation}
\end{Lemma}

\begin{proof}
Let $\epsilon_1=\epsilon/m$, given the Laplace distribution, we have
\begin{center}
$Pr[|\tilde{N}_i| > \epsilon_1]$ = $2 \int_{\epsilon_1}^{\infty} \frac{1}{2b}exp(-\frac{x}{b})$ = $e^{-\epsilon_1 /b}$
\end{center}
then
\begin{center}
$Pr[|\tilde{N}_i| \leq\epsilon_1]$ = 1 - $Pr[|\tilde{N}_i|>\epsilon_1]$ = $ 1 - e^{-\epsilon_1 /b}$
\end{center}
If each $|\tilde{N}_i| \leq\epsilon_1$, we have
$\sum_{i=1}^m |\tilde{N}_i| \leq m \cdot \epsilon_1=\epsilon$, so we have
\begin{displaymath}
Pr[\sum_{i=1}^m |\tilde{N}_i| \leq \epsilon]\geq Pr[|\tilde{N}_i| \leq\epsilon_1]^{m} = (1-e^{-\epsilon_1/b})^{m}
\end{displaymath}
Let $F(x)=(1-x)^m +m x-1$.  The derivative of $F(x)$ is $F'(x)=-m (1-x)^{m -1}+m=m(1-(1-x)^{m -1}) \ge 0$ when $0<x<1$. Note that
$0<e^{-\epsilon_1/b}<1$, so $F(e^{-\epsilon_1/b})\ge F(0) = 0$.
We get
\begin{displaymath}
(1-e^{-\epsilon_1/b})^m \ge 1-m \cdot e^{-\epsilon_1/b}
\end{displaymath}
Recall $\epsilon_1=\epsilon/m$, we derive equation (\ref{equation-Lemma-SumLap}).
\end{proof}

\begin{theorem1}
\label{Theo-utility-Lap} The released $\hat{D}$ of algorithm
\ref{alg-interval} maintains $(\epsilon,\delta)$-usefulness for linear
counting queries, if $\alpha \geq \frac{m \cdot ln(\frac{m}{\delta})}{\epsilon}$, where $m$ is the number of cells in the data cube.
\end{theorem1}

\begin{proof}
Given original data $D$ represented as count vector $\textbf{x}$, using the cell partitioning with Laplace mechanism, the released data $\hat{D}$ can be represented as $\textbf{y} = \textbf{x} + \tilde{\textbf{N}}$, where $\tilde{\textbf{N}}$ is a length-$m$ column vector of Laplace noises drawn from $Lap(b)$ with $b=1/\alpha$.

Given a linear counting query $Q$ represented as $\textbf{q}$ with query size $s$ ($s \le m$), we have
$Q(D)= \textbf{q}\textbf{x}$, and
\begin{displaymath}
Q(\hat{D})= \textbf{Q}\textbf{y} = \textbf{Q}\textbf{x} + \textbf{Q}\tilde{\textbf{N}} = \textbf{Q}\textbf{x} + \sum_{i=1}^s |\tilde{N}_i|
\end{displaymath}

With Lemma \ref{Lemma-SumLap}, we have
\begin{displaymath}
Pr[|Q(D)-Q(\hat{D})| \leq \epsilon] = Pr[\sum_{i=1}^s |\tilde{N}_i| \leq\epsilon] \geq 1-m\cdot exp(-\frac{\epsilon}{ m b})
\end{displaymath}
If $m\cdot exp(-\frac{\epsilon}{m b}) \leq \delta$, then
\begin{displaymath}
Pr[|Q(x,D)-Q(x,\hat{D})| \leq \epsilon]\geq 1-\delta
\end{displaymath}
In order for $m\cdot exp(-\frac{\epsilon}{ m b}) \leq \delta$,
given $b=1/\alpha$, we derive the condition
$\alpha \geq \frac{m \cdot ln(\frac{m}{\delta})}{\epsilon}$.
\end{proof}

\subsection{DPCube: Two-Phase Partitioning}
\vspace{-0.2cm}
\begin{figure}[h!]
\centering
\includegraphics[width=9cm]{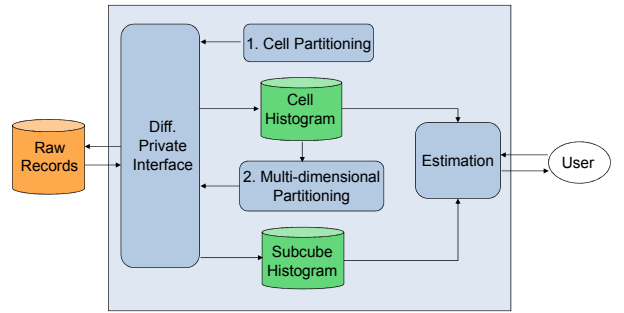}
\vspace{-2.7cm}
\caption{DPCube: 2-phase partitioning}
\label{Figure-release}
\end{figure}

We now present our DPCube algorithm.
DPCube uses an innovative two-phase partitioning strategy as shown in Figure \ref{Figure-release}.
First, a cell based partitioning based on the domains (not the data) is used to generate a fine-grained equi-width cell histogram (as in the baseline strategy), which gives an approximation of the original data distribution.  We generate a synthetic database $D_c$ based on the cell histogram.  Second, a multi-dimensional kd-tree based partitioning is performed on $D_c$ to obtain uniform or close to uniform partitions.  The resulting partitioning keys are used to {\bf Partition} the original database and obtain a {\bf NoisyCount} for each of the partitions, which results in a v-optimal histogram \cite{Poosala:1996}.  Finally, given a user-issued query, an estimation component uses either the v-optimal histogram or both histograms to compute an answer.  The key innovation of our algorithm is that it is data-aware or adaptive because the multi-dimensional kd-tree based partitioning is based on the cell-histogram from the first phase and hence exploits the underlying data distribution indirectly observed by the perturbed cell-histogram.  Essentially, the kd-tree is based on an approximate distribution of the original data. The original database is not queried during the kd-tree construction which saves the privacy budget.  The overall privacy budget is efficiently used and divided between the two phases only for querying the NoisyCount for cells and partitions.  Algorithm \ref{alg-interval2} presents a sketch of the algorithm.

The key step in Algorithm \ref{alg-interval2} is the multi-dimensional partitioning step. It uses a kd-tree based space partitioning strategy that seeks to produce close to uniform partitions in order to minimize the estimation error within a partition.
It starts from the root node which covers the entire space. At each step, a splitting dimension and a split value from the range of the current partition on that dimension are chosen heuristically to divide the space into subspaces.  The algorithm repeats until a pre-defined requirement are met.
In contrast to traditional kd-tree index construction which desires a balanced tree, our main design goal is to generate uniform or close to uniform partitions so that the approximation error within the partitions is minimized.  Thus we propose a uniformity based heuristic to make the decision whether to split the current partition and to select the best splitting point.  There are several metrics that can be used to measure the uniformity of a partition such as information entropy and variance.  In our experiments, we use variance as the metric.  Concretely, we do not split a partition if its variance is greater than a threshold, i.e. it is close to uniform, and split it otherwise.  In order to select the best splitting point, we choose the dimension with the largest range and the splitting point that minimizes the accumulative weighted variance of resulting partitions.  This heuristic is consistent with the goal of a v-optimal histogram which places the histogram bucket boundaries to minimize the cumulative weighted variance of the buckets.
Algorithm \ref{alg-kd} describes the step 2 of Algorithm \ref{alg-interval2}.



\begin{algorithm}
\caption{2-phase partitioning algorithm}
\begin{algorithmic}
\label{alg-interval2}
\REQUIRE{%
$\beta$: number of cells;

$\alpha$: the overall privacy budget}
\STATE
{\bf Phase I:}\\
1. {\bf Partition} the original database based on all domains.\\
2. get {\bf NoisyCount} of each partition using privacy parameter $\alpha_1$ and generate a synthetic dataset $D_c$.\\
{\bf Phase II:}\\
3. Partition $D_c$ by algorithm \ref{alg-kd}.\\
4. {\bf Partition} the original database based on the partition keys returned from step 3.\\
5. release {\bf NoisyCount} of each partition using privacy parameter $\alpha_2=\alpha-\alpha_1$\\
\end{algorithmic}
\end{algorithm}

\begin{algorithm}
\caption{Kd-tree based v-optimal partitioning}
\begin{algorithmic}
\label{alg-kd}
\REQUIRE{$D_t$: input database;
$\xi_0$: variance threshold;
}
\IF{variance of $D_t>\xi_0$}
\STATE Find a dimension and splitting point $m$ which minimizes the cumulative weighted variance of the two resulting partitions;\\
    \STATE Split $D_t$ into $D_{t1}$ and $D_{t2}$ by $m$.\\
            partition $D_{t_1}$ and $D_{t_2}$ by algorithm \ref{alg-kd}.
    \ENDIF
\end{algorithmic}
\end{algorithm}

\partitle{Privacy Guarantee}
We present the theorem below for the 2-phase partitioning algorithm which can be derived directly from the composibility theorems.
\begin{theorem1}
Algorithm \ref{alg-interval2} is $\alpha$-differentially private.
\end{theorem1}
\begin{proof}
Step 2 and Step 5 are $\alpha_1$,$\alpha_2$-differentially private. So the sequence is $\alpha$-differentially private because of theorem \ref{lemma-composition} with $\alpha=\alpha_1+\alpha_2$.
\end{proof}

\partitle{Query Matrix Representation}
We now illustrate how the proposed algorithm can be represented as a query matrix.  We denote $\textbf{H}$ as the query matrix generating the released data in our
algorithm and we have $\textbf{H} = [\textbf{H}_{\textrm{II}};\textbf{H}_{\textrm{I}}]$, where $\textbf{H}_{\textrm{I}}$ and $\textbf{H}_{\textrm{II}}$ correspond to the query matrix in the cell partitioning and kd-tree partitioning phase respectively. $\textbf{H}_{\textrm{I}}$ is an Identity matrix with $m$ rows, each row querying the count of one cell.
$\textbf{H}_{\textrm{II}}$ contains all the partitions generated by the second phase.
We use $\tilde{\textbf{N}}({\alpha})$ to denote the column noise vector and each noise $\tilde{N}_i$ is determined by a differential privacy parameter ($\alpha_1$ in the first phase and $\alpha_2$ in the second phase respectively).
The released data is $\textbf{y} =\textbf{Hx}+\tilde{\textbf{N}}$.  It consists of the cell histogram $\textbf{y}_{\textrm{I}}$ in phase I and the v-optimal histogram $\textbf{y}_{\textrm{I}}$ in phase II: $\textbf{y}_{\textrm{I}}=(\textbf{H}_{\textrm{I}})\textbf{x}+\tilde{\textbf{N}}(\alpha_1)$,
$\textbf{y}_{\textrm{II}}=(\textbf{H}_{\textrm{II}})\textbf{x}+\tilde{\textbf{N}}(\alpha_2)$.

\begin{figure}[h!]
\centering
\vspace{-1cm}
\includegraphics[width=10cm]{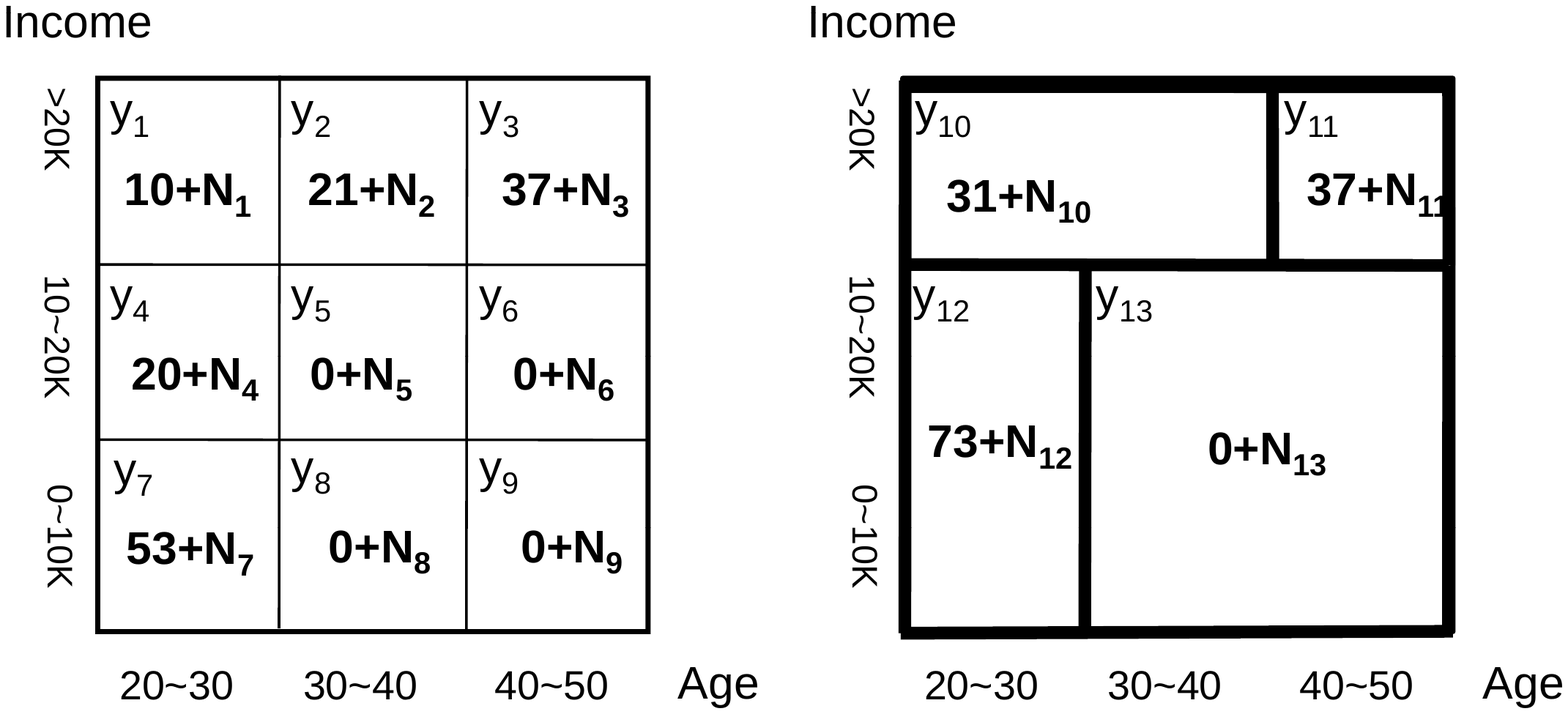}
\vspace{-3cm}
\caption{Example: released cell histogram (left) and subcube histogram (right)}
\label{Fig_exampledata}
\end{figure}

Using our example data from Figure \ref{Fig_exampledata1},
the query matrix $\textbf{H}$ consisting of $\textbf{H}_{\textrm{I}}$ and $\textbf{H}_{\textrm{II}}$ and the released data consisting of the cell histogram $\textbf{y}_{\textrm{I}}$ and subcube histogram $\textbf{y}_{\textrm{II}}$ are shown in Equation (\ref{fml-Identity-matrix}, \ref{fml-P}).  The histograms are also illustrated in Figure \ref{Fig_exampledata}.

\begin{align}
\label{fml-Identity-matrix}
\tiny
\textbf{H}_\textrm{\tiny I}=
\left[
\begin{array}{ccccccccc}
1 & 0 &0 &0 &0 &0&0&0&0\\
0 & 1 &0 &0 &0 &0&0&0&0\\
0 & 0 &1 &0 &0 &0&0&0&0\\
0 & 0 &0 &1 &0 &0&0&0&0\\
0 & 0 &0 &0 &1 &0&0&0&0\\
0 & 0 &0 &0 &0 &1&0&0&0\\
0 & 0 &0 &0 &0 &0&1&0&0\\
0 & 0 &0 &0 &0 &0&0&1&0\\
0 & 0 &0 &0 &0 &0&0&0&1\\
\end{array}
\right]
\textbf{y}_\textrm{\tiny I}=
\left[
\begin{array}{l}
10+\tilde{N}_1(\alpha_1)\\
21+\tilde{N}_2(\alpha_1)\\
37+\tilde{N}_3(\alpha_1)\\
20+\tilde{N}_4(\alpha_1)\\
0+\tilde{N}_5(\alpha_1)\\
0+\tilde{N}_6(\alpha_1)\\
53+\tilde{N}_7(\alpha_1)\\
0+\tilde{N}_8(\alpha_1)\\
0+\tilde{N}_9(\alpha_1)\\
\end{array}
\right]
\\
\label{fml-P}
\tiny
\textbf{H}_{\textrm{\tiny II}}=
\left[
\begin{array}{ccccccccc}
1 & 1 &0 &0 &0 &0&0&0&0\\
0 & 0 &1 &0 &0 &0&0&0&0\\
0 & 0 &0 &1 &0 &0&1&0&0\\
0 & 0 &0 &0 &1 &1&0&1&1
\end{array}
\right]
\textbf{y}_{\textrm{\tiny II}}=
\left[
\begin{array}{l}
51+\tilde{N}_{10}(\alpha_2)\\
37+\tilde{N}_{11}(\alpha_2)\\
53+\tilde{N}_{12}(\alpha_2)\\
0+\tilde{N}_{13}(\alpha_2)
\end{array}
\right]
\end{align}


\subsection{Query Estimation and Error Quantification}
\label{sec-estimation}
Once the histograms are released, given a random user query, the estimation component will compute an answer using the released histograms.  We study two techniques and formally quantify and compare the errors of the query answer they generate.  The first one estimates the query answer using only the subcube histogram assuming a uniform distribution within each partition.  As an alternative approach, we adopt the least squares (LS) method, also used in \cite{Optimizing-PODS,boost-accuracy}, to our two-phase strategy.  The basic idea is to use both cell histogram and subcube histogram and find an approximate solution of the cell counts to resolve the inconsistency between the two histograms.

\partitle{Uniform Estimation using Subcube Histogram}
Based on our design goal to obtain uniform partitions for the subcube histogram, we make the {\em uniform distribution assumption} in the subcube histogram, where the counts in each cell within a partition are assumed to be the same.  Given a partition $p$, we denote $n_p$ as the size of the partition in number of cells.  Hence the count of each cell within this partition is $y_p/n_p$ where $y_p$ is the noisy count of the partition. We denote $\hat{\textbf{x}}_H$ as the estimated cell counts of the original data.  If a query predicate falls within one single partition, the estimated answer for that query is $\frac{s}{n_p}y_p$ where $s$ is the query range size.  Given a random linear query that spans multiple partitions, we can add up the estimated counts of all the partitions within the query predicate. In the rest of the error analysis, we only consider queries within one partition as the result can be easily extended for the general case.

\begin{figure}[h!]
\centering
\vspace{-0.8cm}
\includegraphics[width=9.5cm]{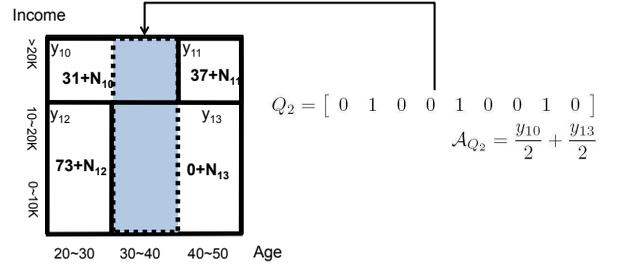}
\vspace{-3.6cm} \caption{Example: query estimation with uniform estimation using subcube histogram}
\label{Fig_estimation_illu}
\end{figure}

Figure \ref{Fig_estimation_illu} shows an example.  Given a query $Q_2$ on population count with age $= [30, 40]$, the original answer is $x_2 + x_5 + x_8$.  Using the subcube histogram, the query overlaps with partition $y_{10}$ and partition $y_{13}$.  Using the uniform estimation, the estimated answer is $\frac{y_{10}}{2} + \frac{y_{13}}{2}$.

\partitle{Error Quantification for Uniform Estimation}
We derive the $(\epsilon,\delta)$-usefulness and the expected error for the uniform estimation method. To formally understand the distribution properties of input database that guarantees the optimality of the method, we first define the smoothness of the distribution within a partition or database similar to \cite{Hardt-smooth}.  The difference is that \cite{Hardt-smooth} defines only the upper bound, while our definition below defines the bound of differences of all cell counts.
Intuitively, the smaller the difference, the smoother the distribution.
\begin{definition1}[$\gamma$-smoothness]
Denote $\textbf{x}$ the original counts of cells in one partition, $\forall x_i,x_j\in \textbf{x}$, if
$|x_j-x_i|\leq \gamma$, then the distribution in the partition satisfies $\gamma$-smoothness.
\end{definition1}

We now present a theorem that formally analyzes the utility of the released data if the input database satisfies $\gamma$-smoothness, followed by a theorem for the general case.

\begin{theorem1}
\label{theo-error-H-uniform}
For $\gamma$-smooth data $\textbf{x}$, given a query $q$ with size $s$, $n_p$ the size of the partition, the uniform estimation method is $(\epsilon,\delta)$-useful if equation (\ref{eqn_usefluness-kd}) holds;
the upper bound of the expected absolute error $\mathbb{E}(\epsilon_H)$ is shown in equation (\ref{eqn_max_E_epsilon_H}).
\begin{align}
\small
\label{eqn_usefluness-kd}
\gamma \leq (\epsilon+\frac{s \cdot log\delta}{\alpha_2 n_p})/min(s,n_p-s)
\\
\small
\label{eqn_max_E_epsilon_H}
\mathbb{E}(\epsilon_H)\leq \gamma [min(s,n_p-s)]+\frac{s}{\alpha_2 n_p}
\end{align}
\end{theorem1}
\begin{proof}
Given a query vector $\textbf{Q}$, the answer using the original data is $\textbf{Qx}$, the estimated answer using the released partition is $\frac{s}{n_p} y_p$ where $y_p$ is the released partition count, $n_p$ is the partition size, and $s$ is the query size.   The released count $y_p=\sum_{i=1}^{n_p}x_i+\tilde{N}(\alpha_2)$.  So we have the absolute error as
\begin{displaymath}
\epsilon_H = |\frac{s}{n_p} y_p-\textbf{Qx}| = |(\frac{s}{n_p}\sum_{i=1}^{n_p}x_i-\textbf{Qx})+\frac{s}{n_p}\tilde{N}(\alpha_2)|
\end{displaymath}
According to the definition of $\gamma$-smoothness, $\forall i,j, |x_j-x_i|\leq \gamma$, then $|\frac{s}{n_p}\sum_{i=1}^{n_p}x_i-\textbf{Qx}| \leq min(s,n_p-s){\gamma}$.
Because of the symmetry of the PDF of $\tilde{N}(\alpha_2)$, we have
\begin{align*}
\label{eqn-e-H-small}
\epsilon_H\leq |min(s,n_p-s){\gamma}+\frac{s}{n_p}\tilde{N}(\alpha_2)|
\\
\leq min(s,n_p-s){\gamma}+|\frac{s}{n_p}\tilde{N}(\alpha_2)|
\end{align*}
By Lemma \ref{lemma-PDF-aZ}, we know the PDF of $\frac{s}{n_p}\tilde{N}(\alpha_2)$.  To satisfy $(\epsilon,\delta)$-usefulness, we require $Pr(\epsilon_H\leq \epsilon)\geq 1-\delta$, then
we derive the condition as in equation (\ref{eqn_usefluness-kd}).

The expected absolute error is
\begin{multline*}
\mathbb{E}(\epsilon_H)= \frac{s}{n_p}\sum_{i=1}^{n_p}x_i-\textbf{Qx}+\frac{s}{n_p}\mathbb{E}|\tilde{N}(\alpha_2)|\\
\leq min(s,n_p-s){\gamma}+\frac{s}{n_p}\mathbb{E}|\tilde{N}(\alpha_2)|
\end{multline*}
Based on the PDF of $\frac{s}{n_p}\tilde{N}(\alpha_2)$, we have
\begin{displaymath}
\mathbb{E}|\tilde{N}(\alpha_2)|=1/\alpha_2
\end{displaymath}
and hence derive equation (\ref{eqn_max_E_epsilon_H}).
\end{proof}

From theorem \ref{theo-error-H-uniform}, we can conclude that if the input data is smoothly distributed or very sparse, $\gamma$ would be small, the error would be
small. In this case, our algorithm achieves the best result.
In the general case, if we do not know the distribution properties of the input data, we present Theorem \ref{theo-error-H-random} to quantify the error.
\begin{theorem1}
\label{theo-error-H-random}
Given a linear counting query $Q$, the expected absolute error for the uniform estimation method,
$\mathbb{E}(\epsilon_H)$, is a function of ($\alpha_1, s,\eta$) shown in
equation (\ref{eqn-error-H-random}):
\begin{align}
\label{eqn-error-H-random}
\mathbb{E}(\epsilon_H)=\int f_s(z,\alpha_1) |\eta+z|dz
\end{align}
where $\eta=\frac{s}{n_p} y_p-Q\textbf{y}_{\textrm{I}}$, $y_p$ is the released count of the partition in the subcube histogram, $n_p$ the size of the partition, $s$ is the size of the query, and $\textbf{y}_{\textrm{I}}$ is the released counts of the cell histogram.
\end{theorem1}
\begin{proof}
Given a query $Q$, the answer using the original data is $Q\textbf{x}$, the estimated answer using the released partition is $\frac{s}{n_p} y_p$.  The released partition count is $y_p=\sum_{i=1}^{n_p}x_i+\tilde{N}(\alpha_2)$.  The released cell counts for the cell histogram in the first phase is $\textbf{y}_{\textrm{I}} = \textbf{x} + \tilde{\textbf{N}}(\alpha_1)$. So we have
\begin{multline*}
\epsilon_H=|\frac{s}{n_p} y_p-Q\textbf{x}|
=|\frac{s}{n_p} y_p - Q(\textbf{y}_{\textrm{I}}-\tilde{\textbf{N}}(\alpha_1))|\\
=|(\frac{s}{n_p}y_p - Q\textbf{y}_{\textrm{I}})+\sum_{i=1}^{s}\tilde{N}_i(\alpha_1)|
\end{multline*}
Denote $\eta=\frac{s}{n_p}y_p-Q\textbf{y}_{\textrm{I}}$, which is the difference (inconsistency) between the estimated answers using the cell histogram and the subcube histogram, then $\epsilon_H=|\eta+\sum_{i=1}^{s}\tilde{N}_i(\alpha_1)|$.
By equation (\ref{fml-baliteral-gamma}) , we have $\mathbb{E}(\epsilon_H)$ in equation (\ref{eqn-error-H-random}).
\end{proof}

\partitle{Least Square Estimation using Cell Histogram and Subcube Histogram}
The least square (LS) method, used in \cite{Optimizing-PODS,boost-accuracy}, finds an approximate (least square) solution of the cell counts that aims to resolve the inconsistency between multiple differentially private views.  In our case, the cell histogram and the subcube histogram provide two views.  We derive the theoretical error of the least square method and compare it with the uniform estimation method. Note that the error quantification in \cite{Optimizing-PODS} is only applicable to the case when $\alpha_1$ and $\alpha_2$ are equal. We derive new result for the general case in which $\alpha_1$ and $\alpha_2$ may have different values.

\begin{theorem1}
\label{theo-E-epsilon-LS}
Given a query $Q$, a least square estimation $\hat{\textbf{x}}_{LS}$ based on the cell histogram and subcube histogram, the expected absolute error of the query answer,
$\mathbb{E}(\epsilon_{LS})$, is a function of $(\alpha_1,\alpha_2,n_p,s)$ in equation (\ref{eqn-error-E-LS}), where $s$ is the size of $Q$, and $n_p$ is the size of the partition.
\begin{multline}
\label{eqn-error-E-LS}
\tiny
\mathbb{E}(\epsilon_{LS})=\frac{(n_p+1)^3}{s^2 (n_p+1-s)} \\\cdot \int |\epsilon| \int f_{n_p-s}(-\frac{(\epsilon-z)(n_p+1)}{s},\alpha_1)\\\cdot\int f_s(\frac{(z-y)(n_p+1)}{n_p+1-s},\alpha_1)f_1(\frac{y(n_p+1)}{s},\alpha_2)dydzd\epsilon
\end{multline}
\end{theorem1}
\begin{proof}
Given our two phase query strategy, the query matrix for the partition is $\textbf{H}=[ones(1,n_p);I_{n_p}]$, where $n_p$ is the partition size.  Using the least square method in \cite{Optimizing-PODS}, we solve $\textbf{H}^{+}=(\textbf{H}^T\textbf{H})^{-1}\textbf{H}^T$, we have
\begin{small}
\begin{align*}
\textbf{H}^{+}=
\left[
\begin{array}{cccccc}
\frac{1}{n_p+1}&\frac{n_p}{n_p+1}&-\frac{1}{n_p+1}&-\frac{1}{n_p+1}&\cdots&-\frac{1}{n_p+1}\\
\frac{1}{n_p+1}&-\frac{1}{n_p+1}&\frac{n_p}{n_p+1}&-\frac{1}{n_p+1}&\cdots&-\frac{1}{n_p+1}\\
\frac{1}{n_p+1}&-\frac{1}{n_p+1}&-\frac{1}{n_p+1}&\frac{n_p}{n_p+1}&\cdots&-\frac{1}{n_p+1}\\
\frac{1}{n_p+1}&-\frac{1}{n_p+1}&-\frac{1}{n_p+1}&-\frac{1}{n_p+1}&\ddots&-\frac{1}{n_p+1}\\
\frac{1}{n_p+1}&-\frac{1}{n_p+1}&-\frac{1}{n_p+1}&-\frac{1}{n_p+1}&\cdots&\frac{n_p}{n_p+1}\\
\end{array}
\right]
\end{align*}
\end{small}
We compute the least square estimation based on the released data $\textbf{y}$ as $\hat{\textbf{x}}_{LS}=\textbf{H}^{+}\textbf{y}$.
So the query answer using the estimation is
\begin{displaymath}
Q\hat{\textbf{x}}_{LS}=Q\textbf{H}^{+}\textbf{y}=Q\textbf{H}^{+}(\textbf{Hx}+\tilde{\textbf{N}}(\alpha))=Q\textbf{x}+Q\textbf{H}^{+}\tilde{\textbf{N}}(\alpha)
\end{displaymath}
The absolute error is
\begin{multline*}
\label{fml-hat-Qx-LS}
Q\hat{\textbf{x}}_{\tiny LS}-Q\textbf{x}=\frac{s}{n_p+1}\tilde{N}(\alpha_2)+\frac{n_p+1-s}{n_p+1}\sum_{i=1}^{s}\tilde{N}(\alpha_1)\\-\frac{s}{n_p+1}\sum_{i=1}^{n_p-s}\tilde{N}(\alpha_1)
\end{multline*}
By Lemma \ref{lemma-PDF-aZ} and equation (\ref{fml-baliteral-gamma}), we know the PDF of $\frac{s}{n_p+1}\tilde{N}(\alpha_2)$, $\frac{n_p+1-s}{n_p+1}\sum_{i=1}^{s}\tilde{N}(\alpha_1)$ and $\frac{s}{n_p+1}\sum_{i=1}^{n_p-s}\tilde{N}(\alpha_1)$, then by convolution formula, we have equation (\ref{eqn-error-E-LS}).
\end{proof}

We will plot the above theoretical results for both uniform estimation and least square estimation with varying parameters in Section \ref{sec-experiment} and demonstrate the benefit of the uniform estimation method, esp. when data is smoothly distributed.

%

\section{Applications}
Having presented the multidimensional partitioning approach for differentially private histogram release, we now briefly discuss the applications that the released histogram can support.

\partitle{OLAP}
On-line analytical processing (OLAP) is a key technology for business-intelligence applications. The computation of multidimensional aggregates, such as count, sum, max, average, is the essence of on-line analytical processing.  The released histograms with the estimation can be used to answer most common OLAP aggregate queries.


\partitle{Classification}
%
Classification is a common data analysis task.  Several recent works studied classification with differential privacy by designing classifier-dependent algorithms (such as decision tree)~\cite{Jagannathan:2009,Friedman:2010}.  The released histograms proposed in this paper can be also used as training data for training a classifier, offering a classifier-independent approach for classification with differential privacy.
To compare the approach with existing solutions~\cite{Jagannathan:2009,Friedman:2010}, we have chosen an ID3 tree classifier algorithm.
However, the histograms can be used as training data for any other classifier.

\partitle{Blocking for Record Linkage}
Private record linkage between datasets owned by distinct parties is another common and challenging problem.  In many situations, uniquely identifying information may not be available and linkage is performed based on matching of other information, such as age, occupation, etc.  Privacy preserving record linkage allows two parties to identify the records that represent the same real world entities without disclosing additional information other than the matching result.
While Secure Multi-party Computation (SMC) protocols can be used to provide strong security and perfect accuracy, it incurs prohibitive computational and communication cost in practice.  Typical SMC protocols require $O(n*m)$ cryptographic operations where $n$ and $m$ correspond to numbers of records in two datasets.  In order to improve the efficiency of record linkage, \textit{blocking} is generally in use according to \cite{Elma07}.  The purpose of blocking is to divide a dataset into mutually exclusive blocks assuming no matches occur across different blocks.  It reduces the number of record pairs to compare for the SMC protocols but meanwhile we need to devise a blocking scheme that provides strong privacy guarantee.  \cite{inan10private} has proposed a hybrid approach with a differentially private blocking step followed by an SMC step.  The differentially private blocking step adopts tree-structured space partitioning techniques and uses Laplace noise at each partitioning step to preserve differential privacy.  The matching blocks are then sent to SMC protocols for further matching of the record pairs within the matching blocks which significantly reduces the total number of pairs that need to be matched by SMC.



The released histograms in this paper can be also used to perform blocking which enjoys differential privacy.
Our experiments show that we can achieve approximately optimal reduction ratio of the pairs that need to be matched by SMC with appropriate setting for the threshold $\xi_0$.  We will examine some empirical results in Section \ref{sec-experiment}.

\section{Experiment}
\label{sec-experiment}
We first simulate and compare the theoretical query error results from Section \ref{sec-estimation} for counting queries that falls within one partition to show the properties and benefits of our approach (Section \ref{sec-theoretical}).
We then present a set of experimental evaluations of the quality of the released histogram in terms of weighted variance (Section \ref{sec-variance}), followed by evaluations of query error against random linear counting queries and a comparison with existing solutions (Section \ref{sec-kd}).
Finally, we also implemented and experimentally evaluate the two additional applications using the released histograms, classification and blocking for record linkage, and show the benefit of our approach (Section \ref{sec-app}).

\subsection{Simulation Plots of Theoretical Results}
\label{sec-theoretical}
As some of the theoretical results in Section \ref{sec-estimation} consists of equations that are difficult to compute for given parameters, we use the simulation approach to show the results and the impact of different parameters in this section.
Detailed and sophisticated techniques about the simulation approach can
be found in \cite{book-statistical-inference} and
\cite{book-probability-models}.  Table \ref{tbl-parameters} shows the parameters used in the simulation experiment and their default values.

\begin{table}[h!]
\label{tbl-parameters}
\caption{Simulation parameters}
\begin{tabular}{|c|c|c|}
\hline
parameter&description &default value\\
\hline
$n_p$ & partition size & $n_p=11$\\
\hline
$s$ & query size & $s\leq n_p$\\
\hline
$\alpha_1$, $\alpha_2$ & diff. privacy parameters & $\alpha_1=0.05$, $\alpha_2=0.15$\\
\hline
$\gamma$ & smoothness parameter & $\gamma=5$\\
\hline
$\eta$ & inconsistency between &$\eta=5$\\
& cell and subcube histogram & \\
\hline
\end{tabular}
\end{table}

\partitle{Metric}
We evaluate the error of absolute count queries. Recall that $\mathbb{E}(\epsilon_{LS})$ is the expected error of the least square estimation method; $max(\mathbb{E}(\epsilon_H))$ is the upper bound of expected error of the uniform estimation method when the data is $\gamma$-smooth; $\mathbb{E}(\epsilon_H)$ is the expected error of the uniform estimation method in the general case.
Note that $\mathbb{E}(\epsilon_{LS})$, $max(\mathbb{E}(\epsilon_H))$, and $\mathbb{E}(\epsilon_H)$ are derived from equation (\ref{eqn-error-E-LS}), (\ref{eqn_max_E_epsilon_H}), (\ref{eqn-error-H-random}) respectively.

\begin{figure}[h!]
\begin{minipage}{0.23\textwidth}
\centering
\includegraphics[width=4.5cm]{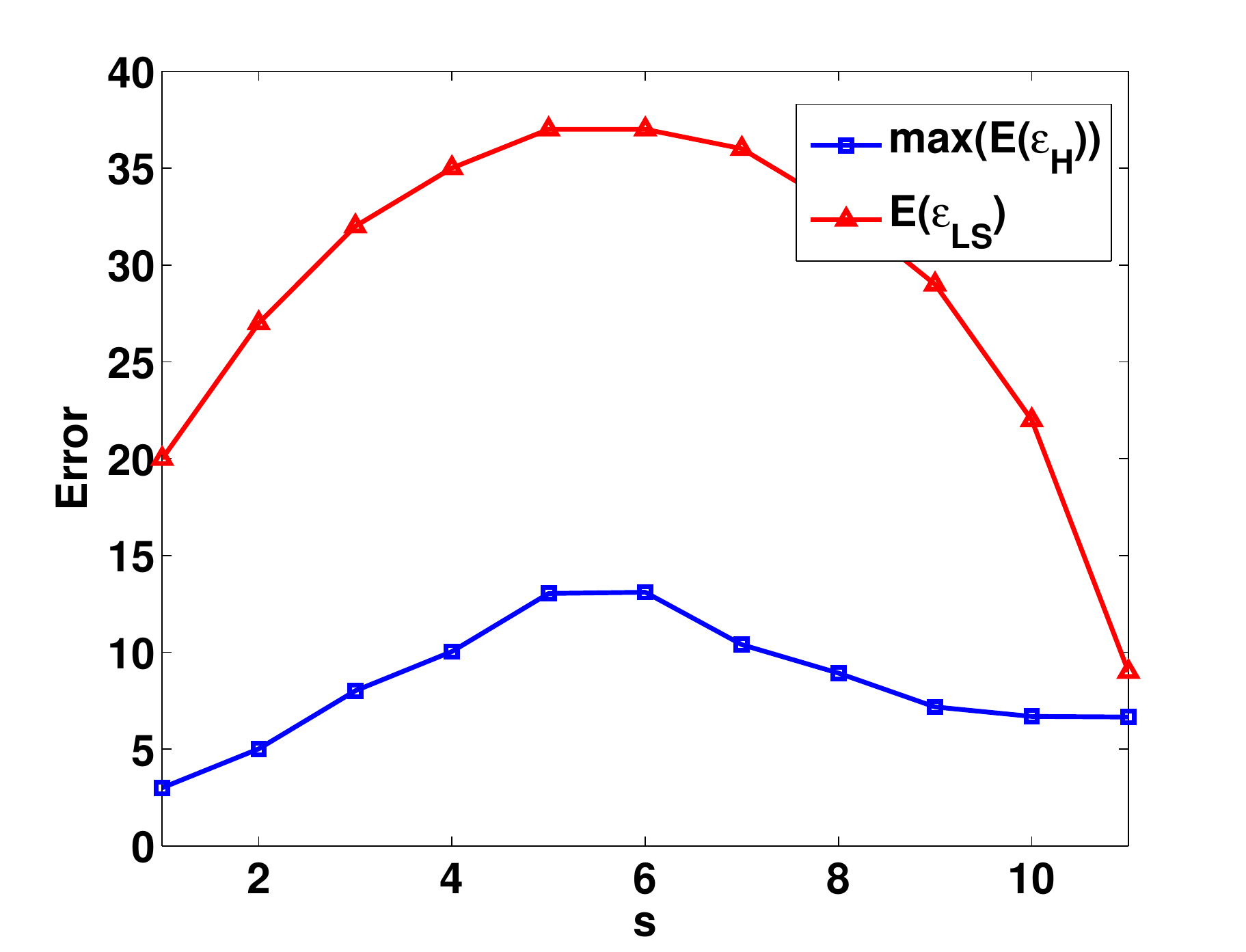}
\mbox{(a) $\gamma$-smooth distribution}
\end{minipage}
\begin{minipage}{0.23\textwidth}
\centering
\includegraphics[width=4.5cm]{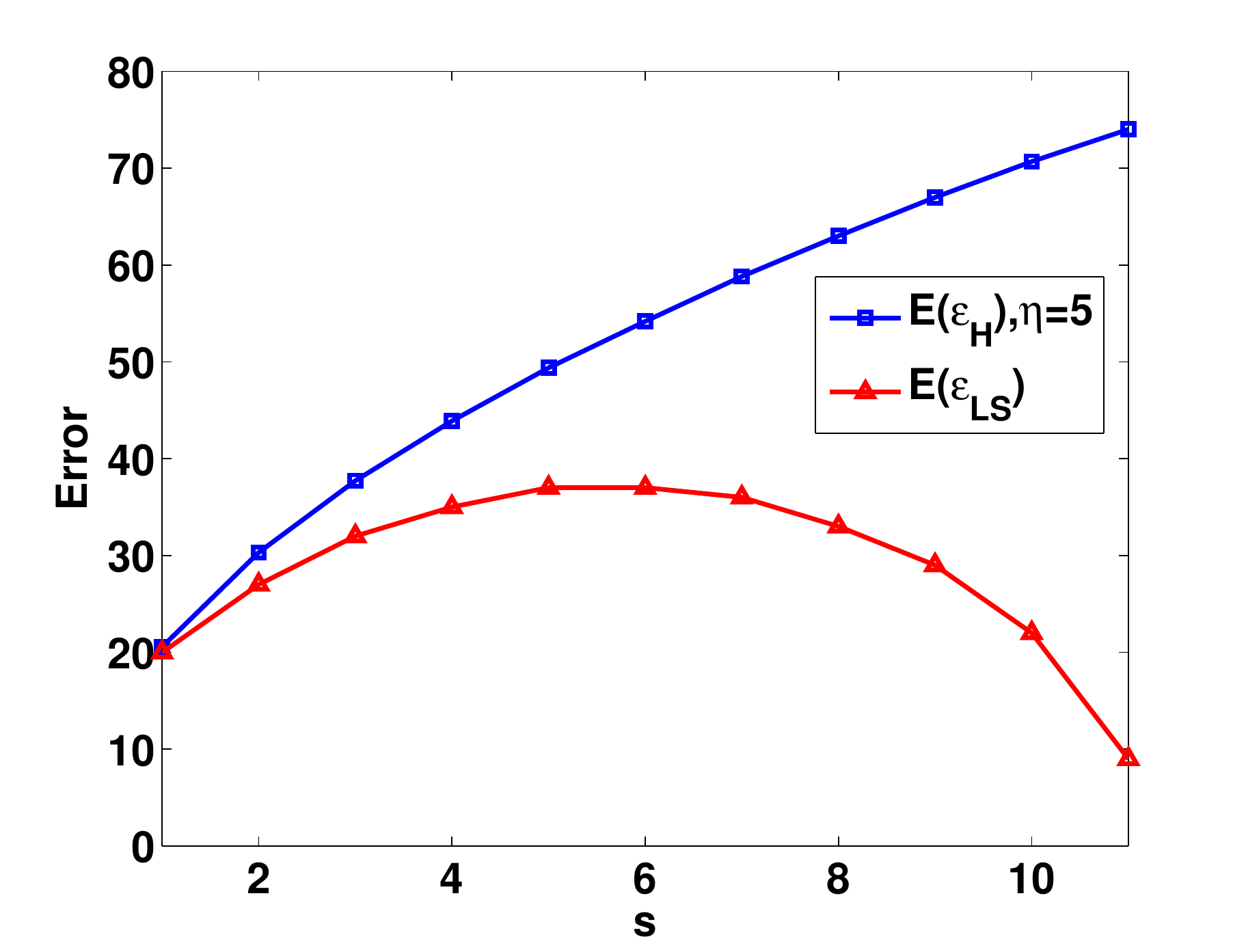}
\mbox{(b) any distribution}
\end{minipage}
\caption{Query error vs. query size $s$}
\label{Figure-E-error-LS-s}
\end{figure}

\partitle{Impact of Query Size}
We first study the impact of the query size.  Figure \ref{Figure-E-error-LS-s} shows the error of the uniform and least square solutions with respect to varying query size for $\gamma$-smooth data and general case (any distribution) respectively.
We can see that the highest error appears when the query size $s$
is half of the partition size $n_p$. When $\gamma$ is small, i.e. data is smoothly distributed, the uniform estimation method outperforms the least square method.  In the general case when we do not have any domain knowledge about the data distribution, it is beneficial to use both cell histogram and subcube histogram to resolve their inconsistencies.

\begin{figure}[h!]
\begin{minipage}{0.23\textwidth}
\centering
\includegraphics[width=4.5cm]{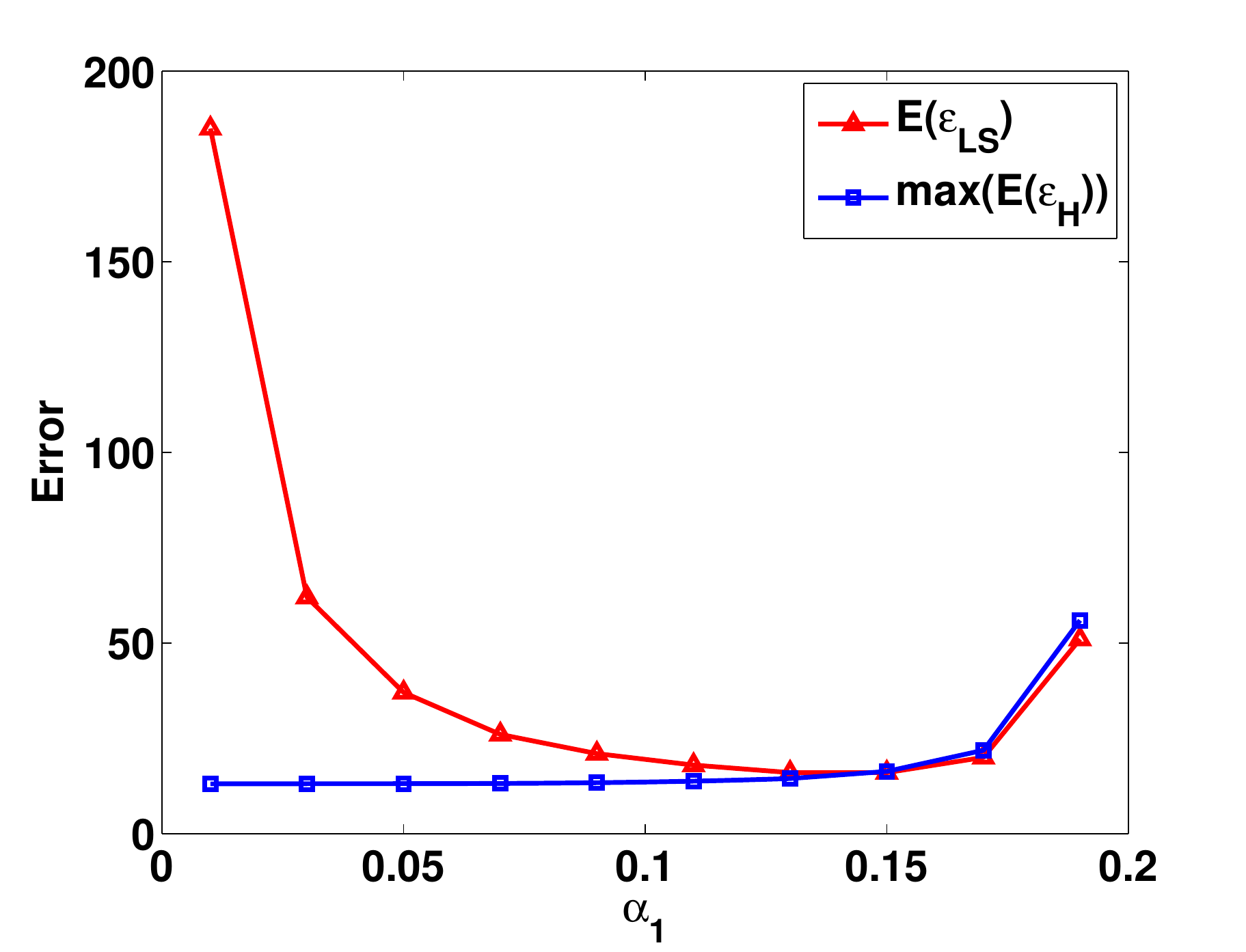}
\mbox{(a) $\gamma$-smooth distribution}
\end{minipage}
\begin{minipage}{0.23\textwidth}
\centering
\includegraphics[width=4.5cm]{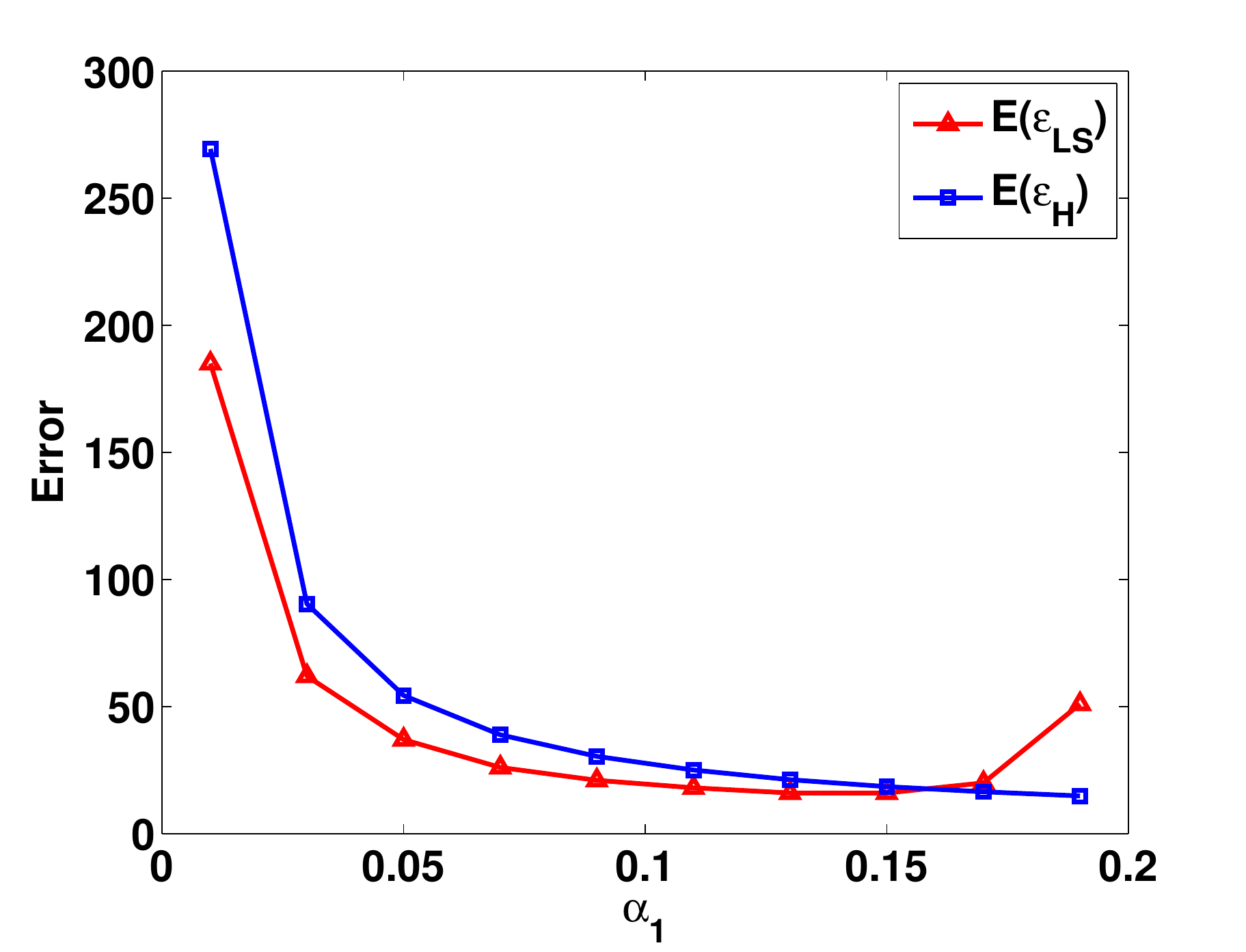}
\mbox{(b) any distribution}
\end{minipage}
\caption{Query error vs. privacy budget allocation $\alpha_1$}
\label{Figure-E-error-LS-alpha1}
\end{figure}

\partitle{Impact of Privacy Budget Allocation}
We now study the impact of the allocation of the overall differential privacy budget $\alpha$ between the two phases.  The overall budget is fixed with $\alpha=0.2$.
Figure \ref{Figure-E-error-LS-alpha1} shows the error of the uniform and least square solutions with respect to varying query size for $\gamma$-smooth data and general case (any distribution) respectively.
For the LS method, equally dividing $\alpha$ between the two phases or slightly more for cell-based partitioning works better than other cases. The results for the uniform estimation method present interesting trends. For smoothly distributed data, a smaller privacy budget for the partitioning phase yields better result. Intuitively, since data is smoothly distributed, it is beneficial to save the privacy budget for the second phase to get a more accurate overall partition count.   On the contrary, for the random case, it is beneficial to spend the privacy budget in the first phase to get a more accurate cell counts, and hence more accurate partitioning.

\begin{figure}[h!]
\begin{minipage}{0.23\textwidth}
\centering
\includegraphics[width=4.5cm]{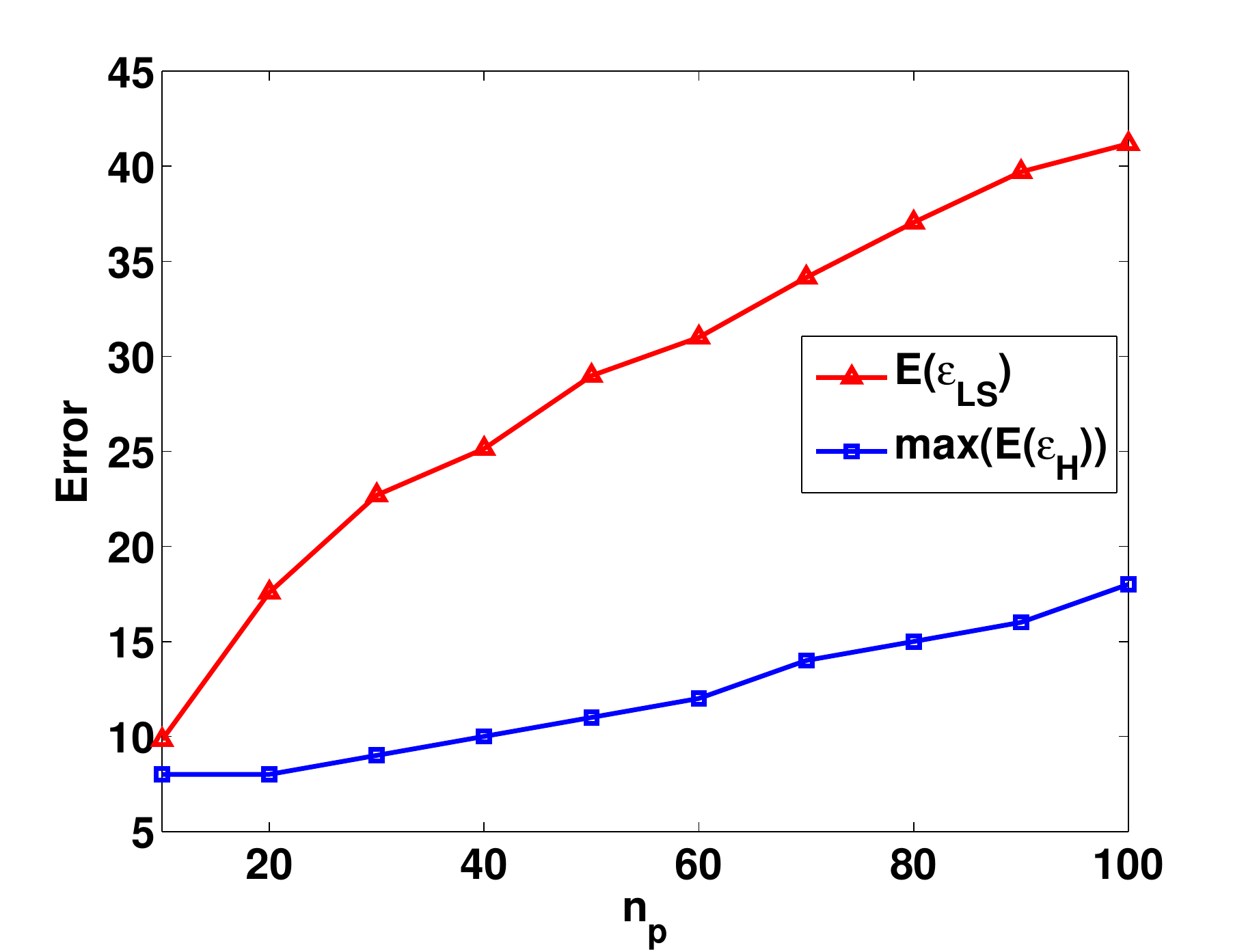}
\mbox{(a) vs. partition size $n_p$}
\end{minipage}
\begin{minipage}{0.23\textwidth}
\centering
\includegraphics[width=4.5cm]{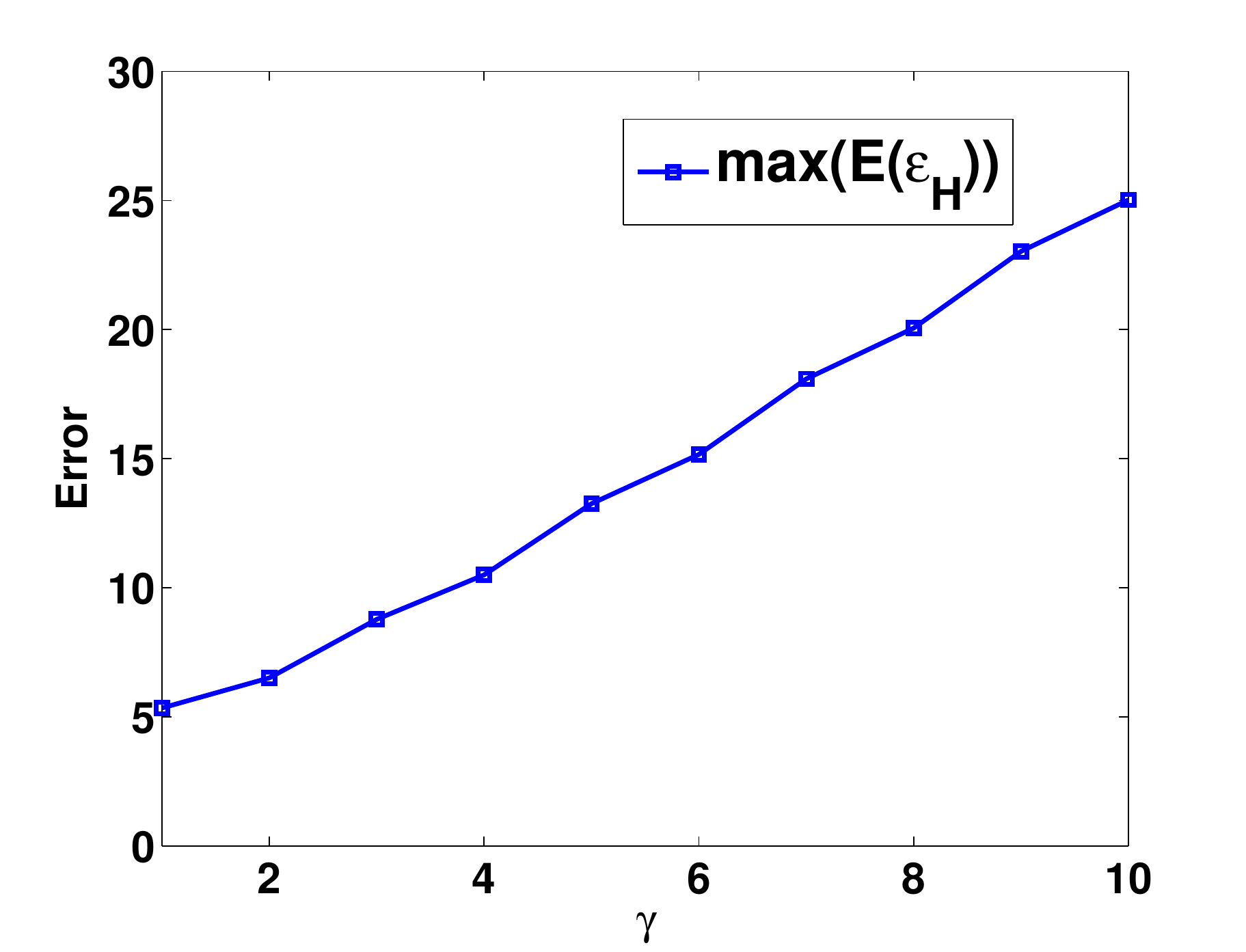}
\mbox{(b) vs. smoothness $\gamma$}
\end{minipage}
\caption{Query error for $\gamma$-smooth distribution}
\label{Figure-E-error-LS-n}
\end{figure}

\partitle{Impact of Partition Size}
For $\gamma$-smooth data, the expected error is dependent on the partition size $n_p$ and the smoothness level $\gamma$.  Figure
\ref{Figure-E-error-LS-n}(a) shows the error of the uniform and least square solutions with respect to varying partition size $n_p$ for $\gamma$-smooth data.
We observe that the error increases when the partition size increases because of the increasing approximation error within the partition. Therefore, a good partitioning algorithm should avoid large partitions.

\partitle{Impact of Data Smoothness}
For $\gamma$-smooth data, we study the impact of the smoothness level $\gamma$ on the error bound for the uniform estimation method.  Figure \ref{Figure-E-error-LS-n}(b) shows the maximum error bound with varying $\gamma$. We can see that the more smooth the data, the less
error for released data.
Note that the query size $s$ is set to be nearly half size of the partition size $n_p$ as default, which magnifies the impact of the smoothness.  We observed in other parameter settings that the error increases only slowly for queries with small or big query sizes.

\begin{figure}[h!]
\centering
\includegraphics[width=4.5cm]{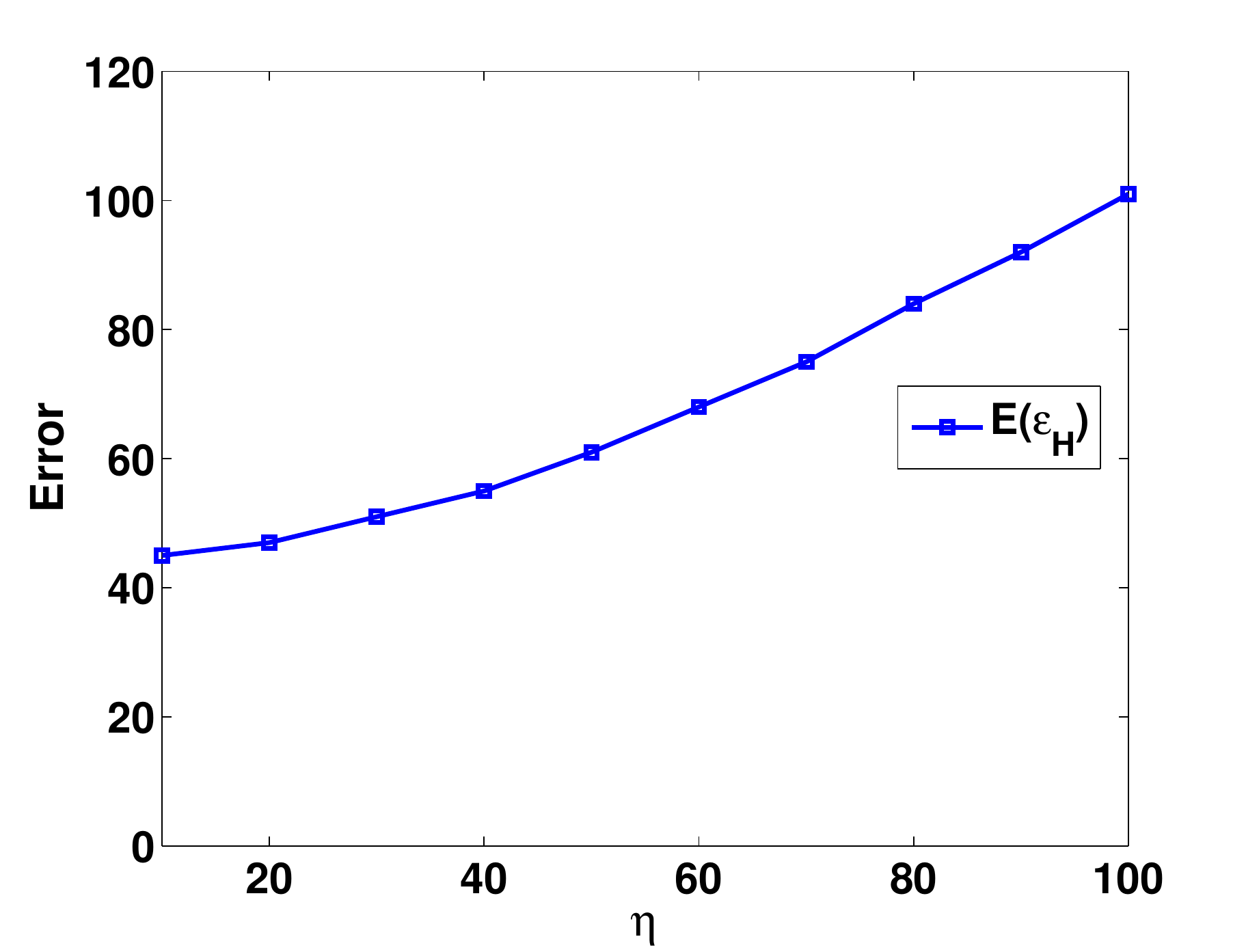}
\caption{Query error vs. level of inconsistency $\eta$ for any distribution}
\label{Figure-E-error-H-Y-eta}
\end{figure}

\partitle{Impact of Inconsistency} For data with any (unknown) distribution, $\mathbb{E}(\epsilon_H)$ is a function of $\eta$, the level of inconsistency between the cell histogram and the subcube histogram.  In contrast to the the $\gamma$-smooth case in which we have a prior knowledge about the smoothness of the data, here we only have this observed level of inconsistency from the released data which reflects the smoothness of the original data.  Figure \ref{Figure-E-error-H-Y-eta} shows the error for the uniform estimation method with varying $\eta$. Note that the error increases with increasing $\eta$ and even when $\eta=10$, the error is still lager than the error of least square method in Figure \ref{Figure-E-error-LS-n}.

\subsection{Histogram Variance}
\label{sec-variance}
We use the Adult dataset from the Census~\cite{Frank:2010}.  All experiments were run on a computer with Intel P8600($2*2.4$ GHz) CPU and 2GB memory.
For computation simplicity and smoothness of data distribution, we only use the first $10^4$ data records.

\begin{figure}[h!]
\begin{minipage}{0.23\textwidth}
\centering
\includegraphics[width=4.5cm]{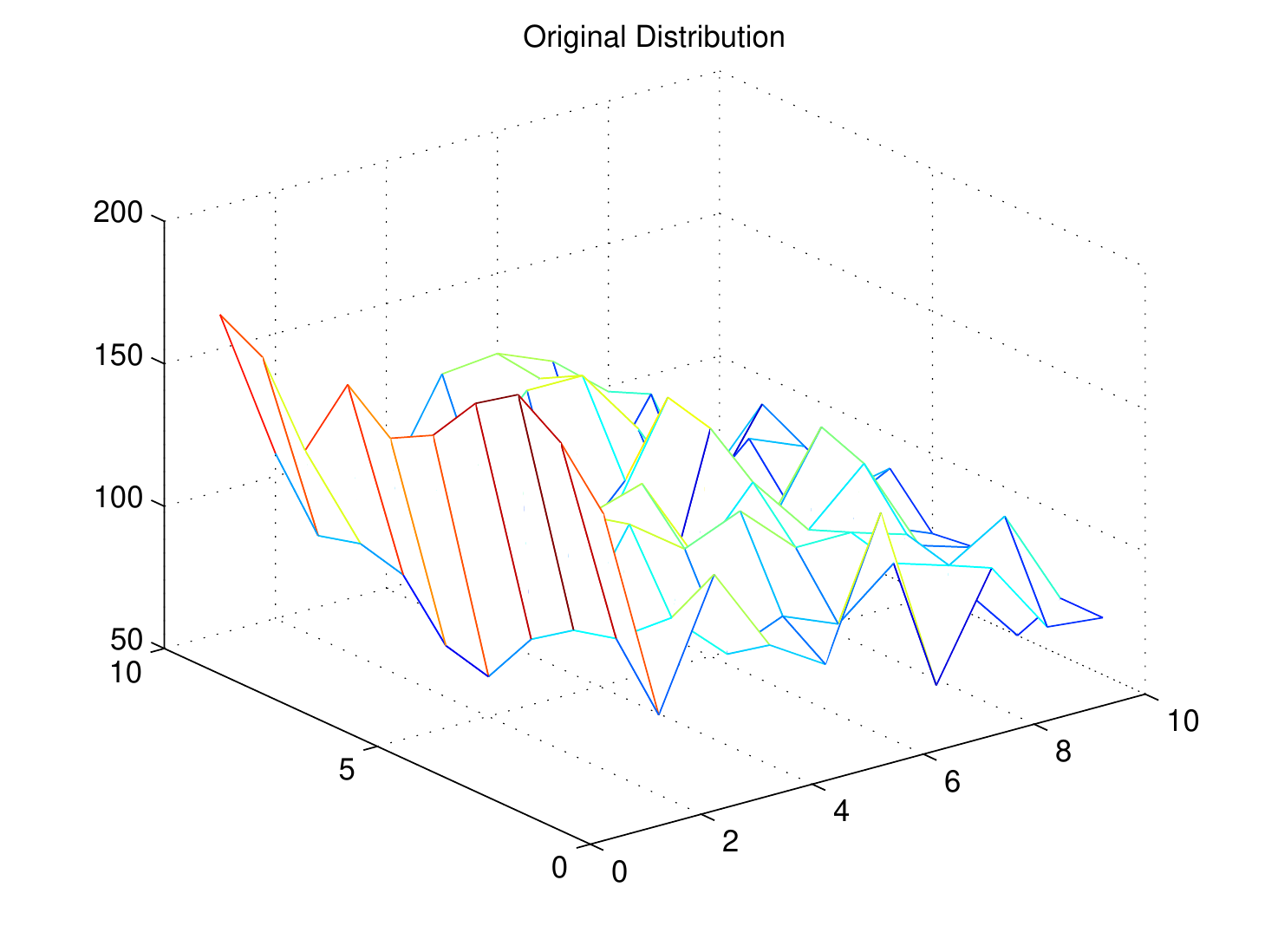}
\mbox{(a) Original histogram}
\end{minipage}
\begin{minipage}{0.23\textwidth}
\centering
\includegraphics[width=4.5cm]{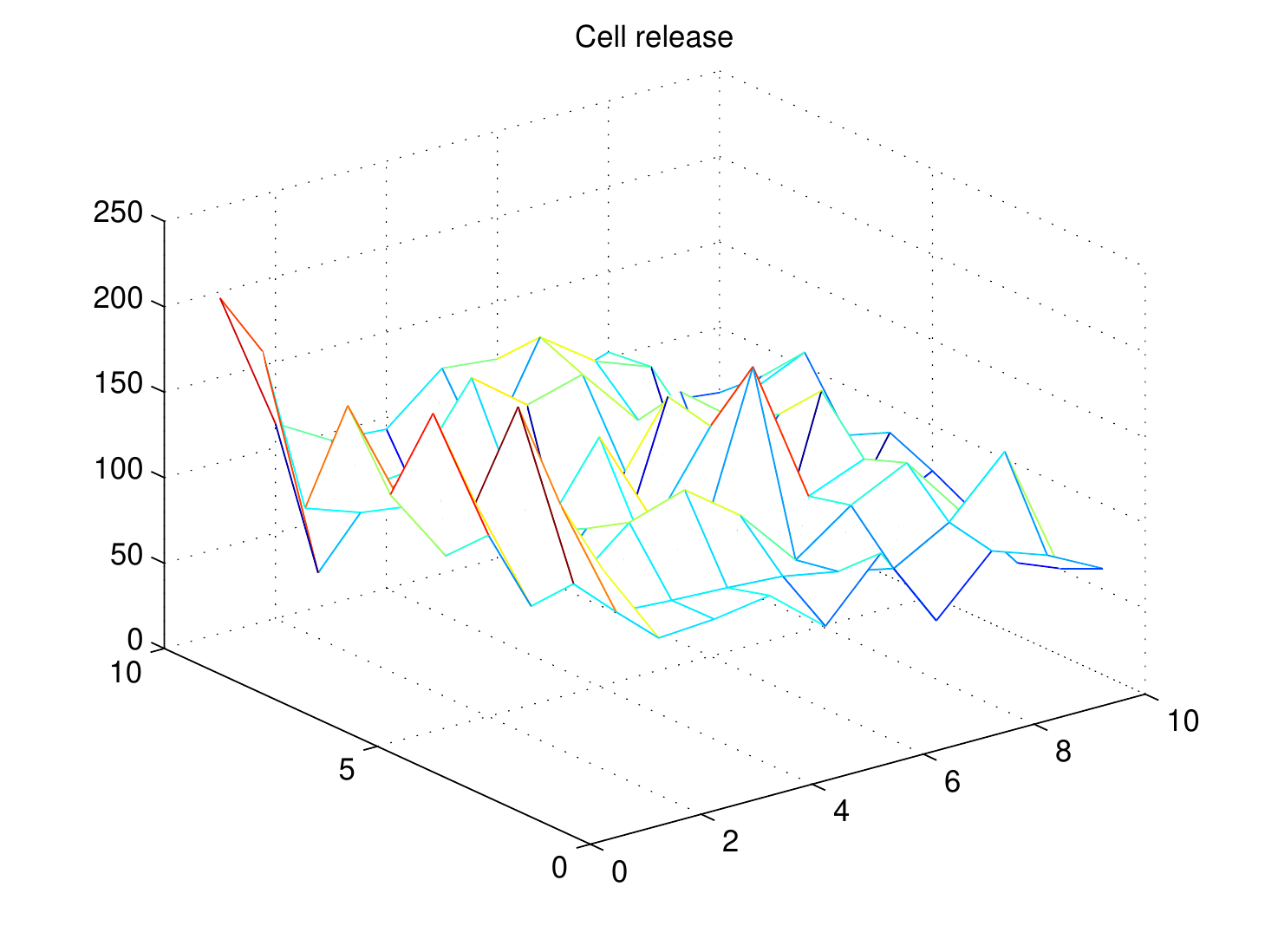}
\mbox{(b) Cell histogram}
\end{minipage}\\
\begin{minipage}{0.23\textwidth}
\centering
\includegraphics[width=4.5cm]{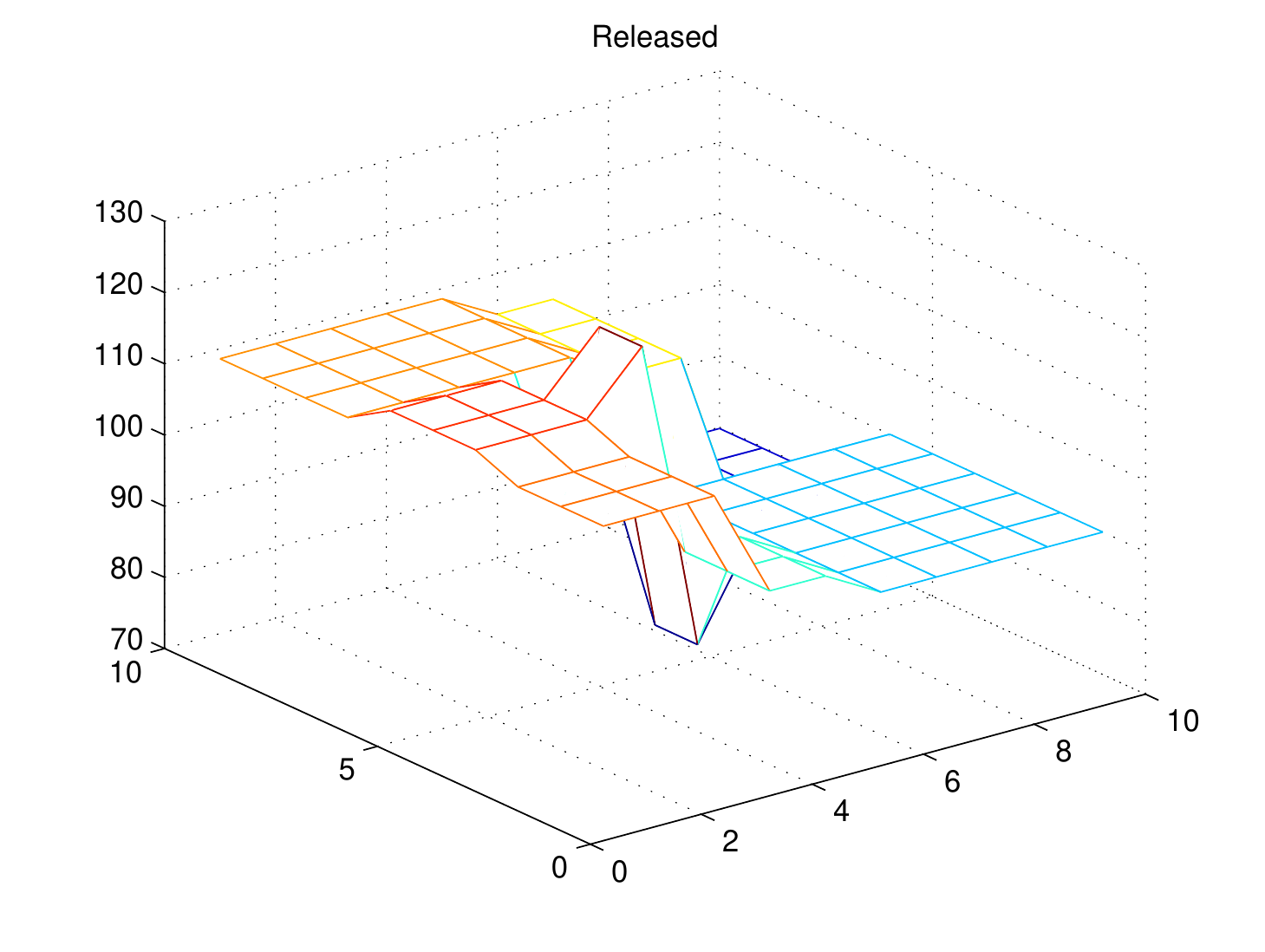}
\mbox{(c) Subcube histogram}
\end{minipage}
\begin{minipage}{0.23\textwidth}
\centering
\includegraphics[width=4.5cm]{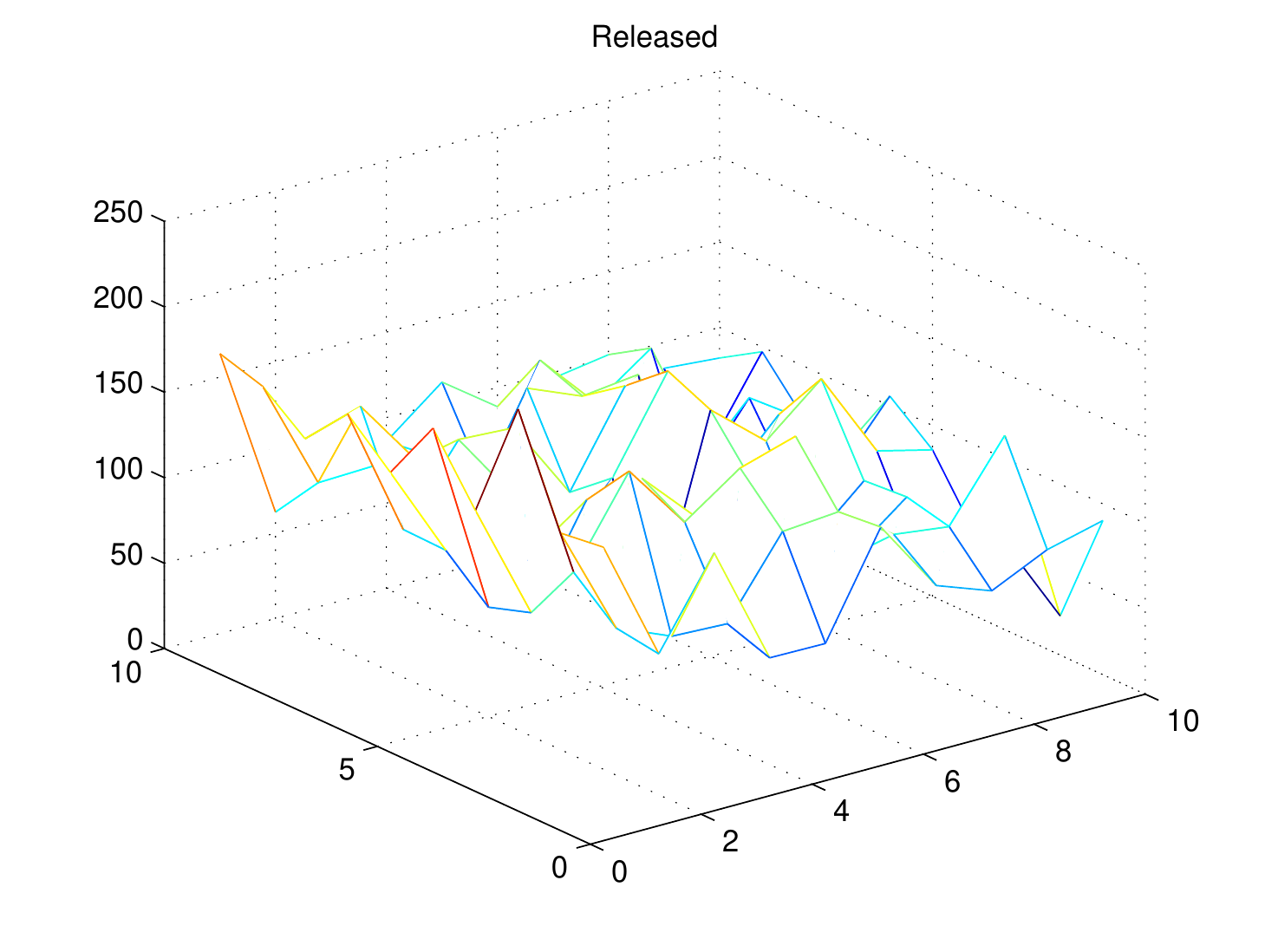}
\mbox{(d) Estimated cell histogram}
\end{minipage}
\caption{Original and released histograms}
\label{Figure-histograms}
\end{figure}

\partitle{Original and Released Histograms}
We first present some example histograms generated by our algorithm for the Census dataset.  Figure \ref{Figure-histograms}(a) shows the original 2-dimensional histogram on Age and Income.  Figure \ref{Figure-histograms}(b) shows a cell histogram generated in the first phase with $\alpha_1=0.05$.  Figure \ref{Figure-histograms}(c) shows a subcube histogram generated in the second phase with estimated cell counts using uniform estimation, in which each horizontal plane represents one partition.  Figure \ref{Figure-histograms}(d) shows an estimated cell histogram using the LS estimation using both cell histogram and subcube histogram.  We systematically evaluate the utility of the released subcube histogram with uniform estimation below.

\partitle{Metric}
We now evaluate the quality of the released histogram using an application-independent metric, the weighted variance of the released subcube histogram.  Ideally, our goal is to obtain a v-optimal histogram which minimizes the weighted variance so as to minimize the estimation error within partitions.   Formally, the weighted variance of a histogram is defined as $V =
\sum_{i=1}^{p}x_iV_i$, where $p$ is the number of
partitions, $x_i$ is the number of data points in the $i$-th
partition, and $V_i$ is the variance in the $i$-th
partition~\cite{Poosala:1996}.


\begin{figure}[h!]
\begin{minipage}{0.23\textwidth}
\centering
\includegraphics[width=4.5cm]{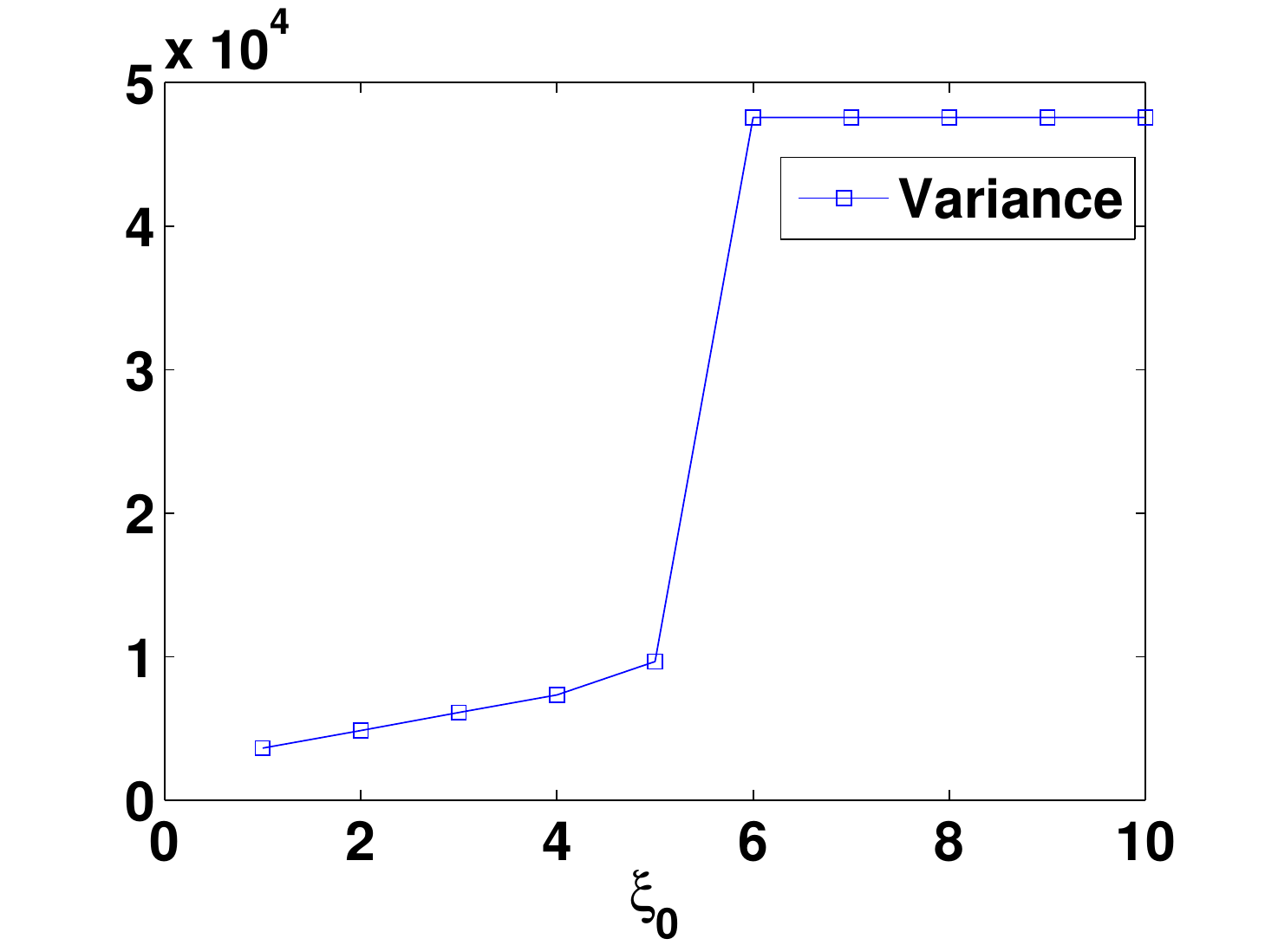}
\mbox{((a) vs. threshold $\xi_0$}
\end{minipage}
\begin{minipage}{0.23\textwidth}
\centering
\includegraphics[width=4.5cm]{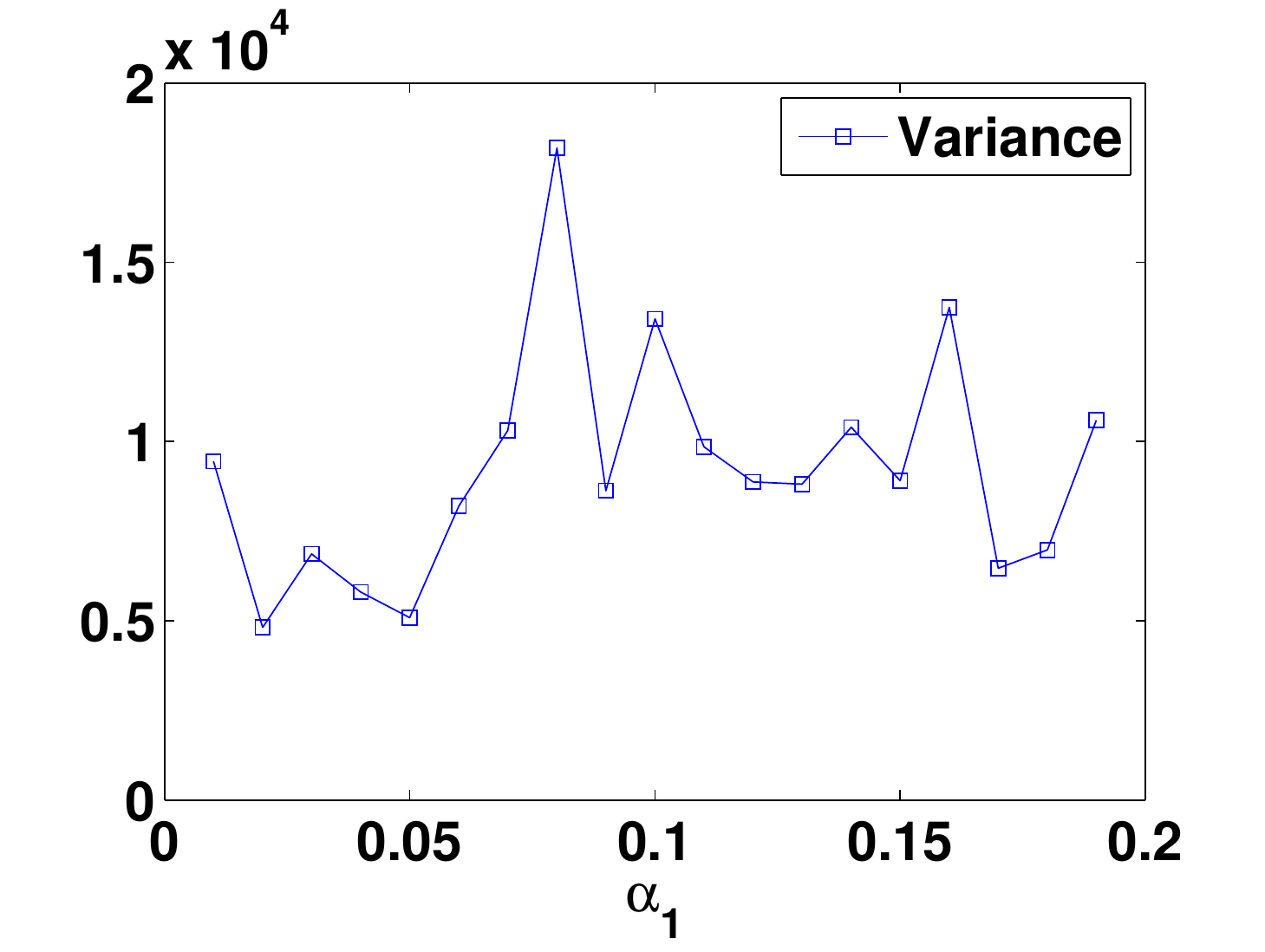}
\mbox{(b) vs. privacy budget $\alpha_1$}
\end{minipage}
\caption{Histogram variance and impact of parameters}
\label{fig:variance}
\end{figure}

\partitle{Impact of Variance Threshold}
Figure~\ref{fig:variance}(a) shows the weighted variance of the released subcube histogram with respect to varying variance threshold value $\xi_0$ used in our algorithm. As expected, when $\xi_0$ becomes large, less partitioning is performed, i.e. more data points are grouped into one bucket, which increases the variance.

\partitle{Impact of Privacy Budget Allocation}
Figure~\ref{fig:variance}(b) shows the weighted variance with respect to varying privacy budget in the first phase $\alpha_1$ (with fixed $\xi_0$. Note that the overall privacy budget is fixed.
We see that the correlation is not very clear due to the randomness of noises. Also the $\xi_0$ is fixed which does not reflect the different magnitude of noise introduced by different $\alpha_1$.
Generally speaking, the variance should decrease gradually when $\alpha_1$ increases because large $\alpha_1$ will introduce less noise to the first phase, which will lead to better partitioning quality.


\subsection{Query Error}
\label{sec-kd}

We evaluate the quality of the released histogram using random linear counting queries and measure the average absolute query error and compare the results with other algorithms.
We use the Age and Income attributes and generated $10^5$ random counting queries to calculate the average query error.  The random queries consist of two range predicates on the two attributes respectively.
We also implemented an alternative kd-tree strategy similar to that used in \cite{inan10private}, referred to as hierarchical kd-tree, and another data release algorithm \cite{boost-accuracy}, referred to as consistency check, for comparison purposes.

\begin{figure}[h!]
\begin{minipage}{0.23\textwidth}
\centering
\includegraphics[width=4.5cm]{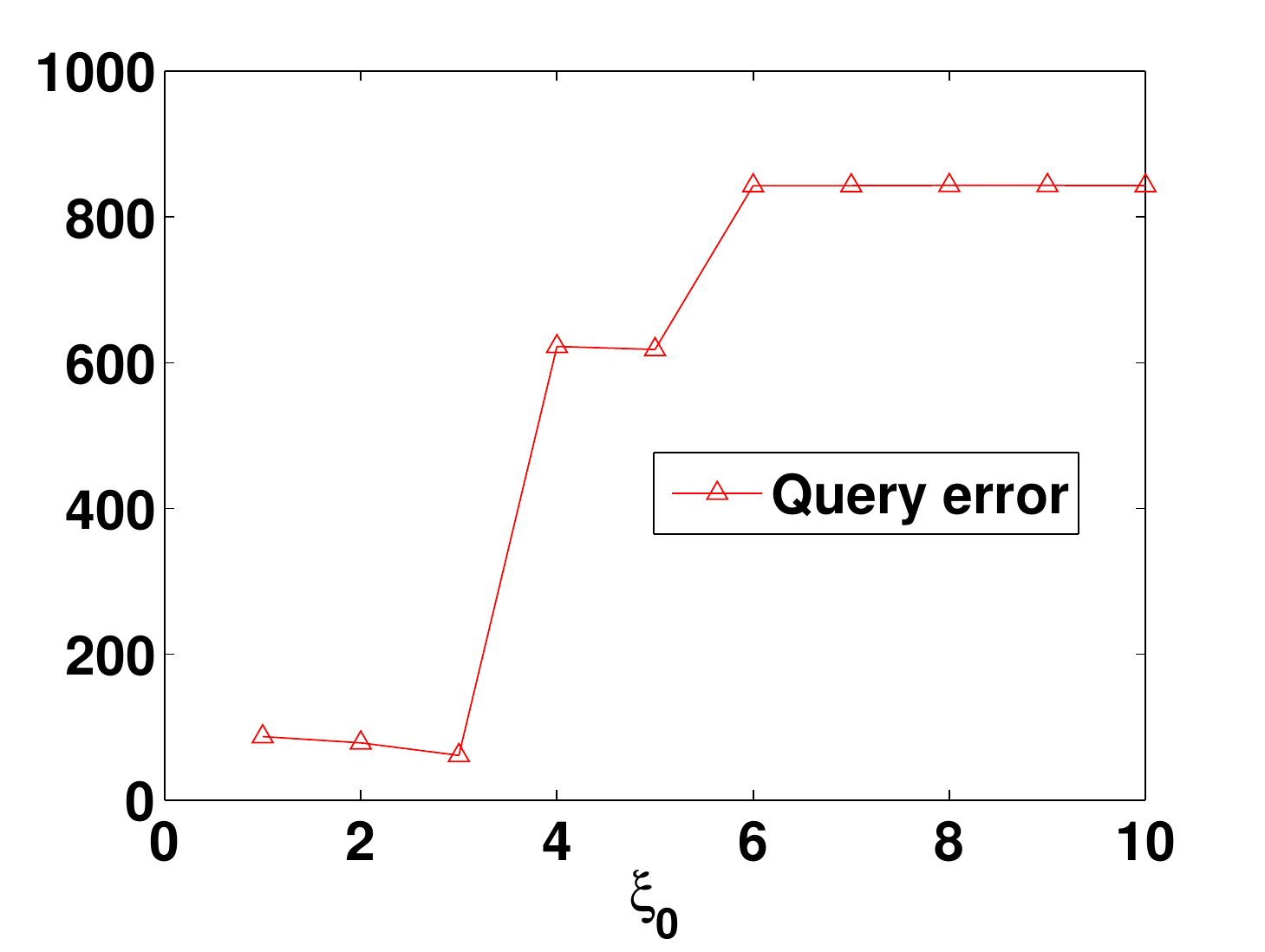}
\mbox{(a) vs. threshold $\xi_0$}
\end{minipage}
\begin{minipage}{0.23\textwidth}
\centering
\includegraphics[width=4.5cm]{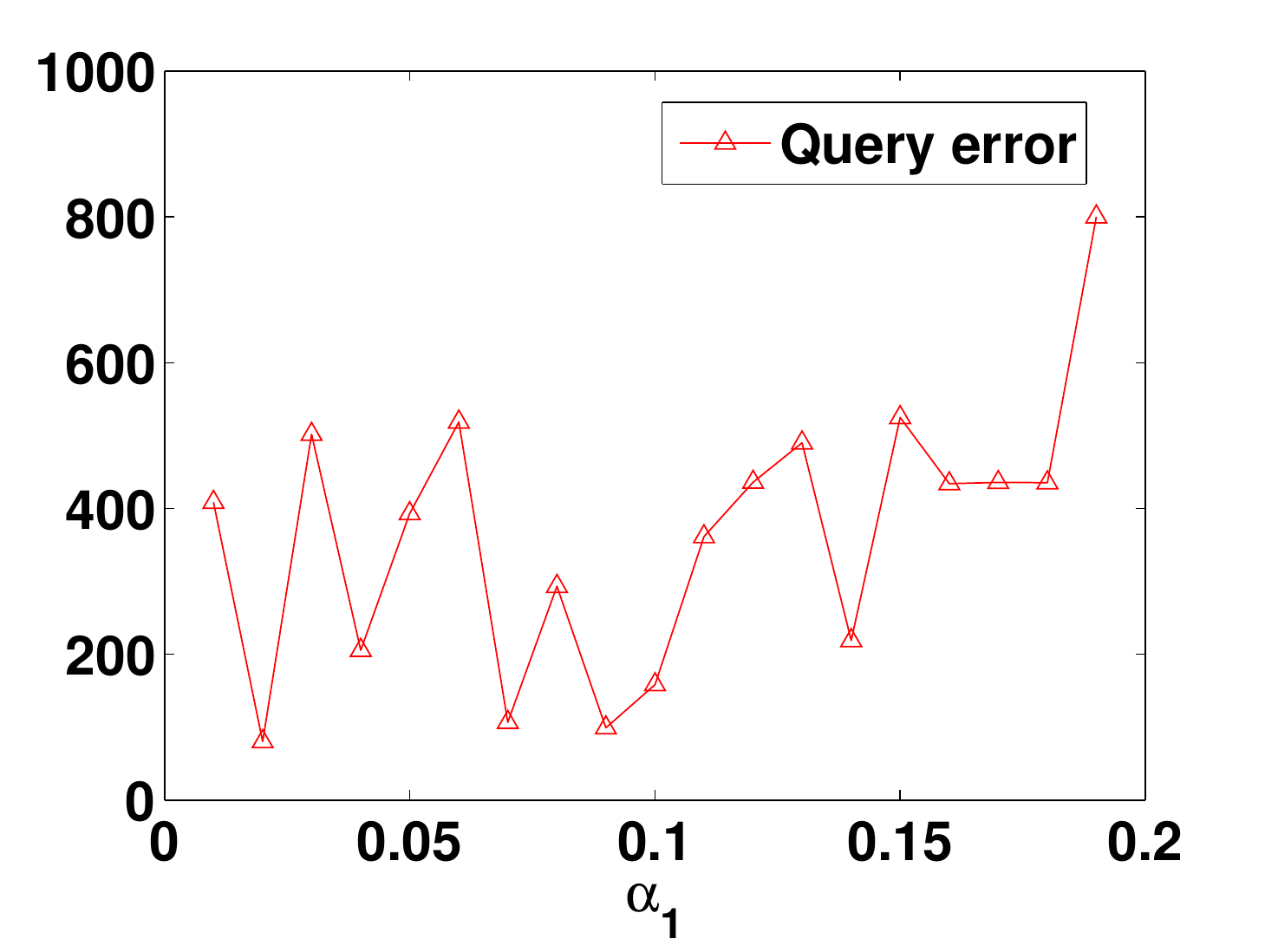}
\mbox{(b) vs. privacy budget $\alpha_1$}
\end{minipage}
\caption{Query error and impact of parameters}
\label{Figure-vsErr_ERRvsThresh}
\end{figure}

\partitle{Impact of Variance Threshold}
We first evaluate the impact of the threshold value $\xi_0$ on query error.  Figure \ref{Figure-vsErr_ERRvsThresh}(a) shows the average absolute query error with respect to varying threshold values.  
Consistent with Figure~\ref{fig:variance}, we observe that the query error increases with an increasing threshold value, due to the increased variance within partitions.

\partitle{Impact of Privacy Budget Allocation}
We next evaluate the impact of how to allocate the overall privacy budget $\alpha$ into $\alpha_1$, $\alpha_2$ for the two phases in the algorithm.
Figure \ref{Figure-vsErr_ERRvsThresh}(b) shows the average absolute query error vs. varying $\alpha_1$.
We observe that small $\alpha_1$ values yield better results, which complies with our theoretical result for $\gamma$-smooth data in figure \ref{Figure-E-error-LS-alpha1}.  This verifies that the real life Adult dataset has a somewhat smooth distribution.
On the other hand, for data with unknown distribution, we expect $\alpha_1$ cannot be too small as it is more important for generating a more accurate partitioning.

\begin{figure}[h!]
\begin{minipage}{0.23\textwidth}
\centering
\includegraphics[width=4.5cm]{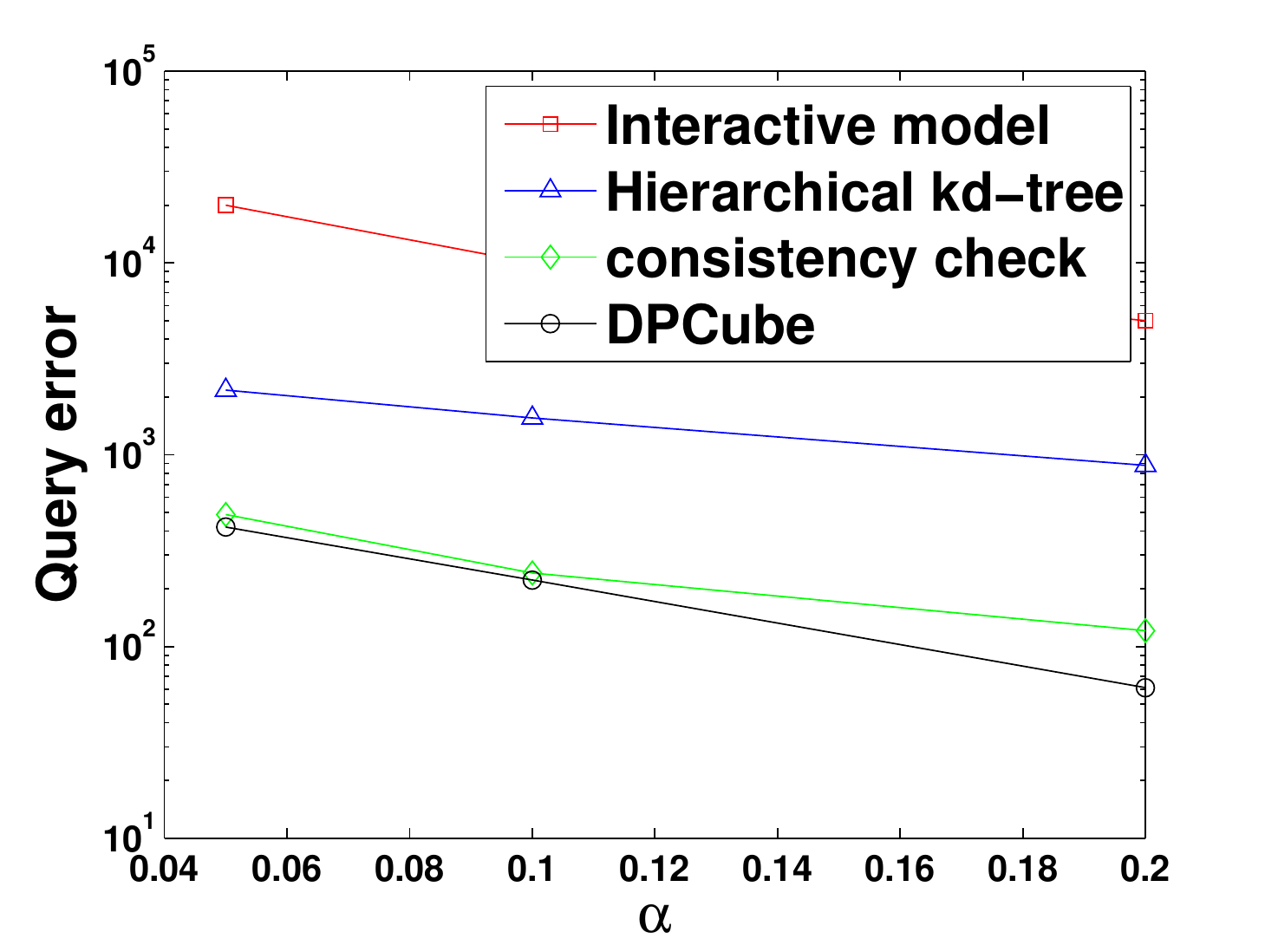}
\mbox{(a) vs. privacy budget $\alpha$}
\end{minipage}
\begin{minipage}{0.23\textwidth}
\centering
\includegraphics[width=4.5cm]{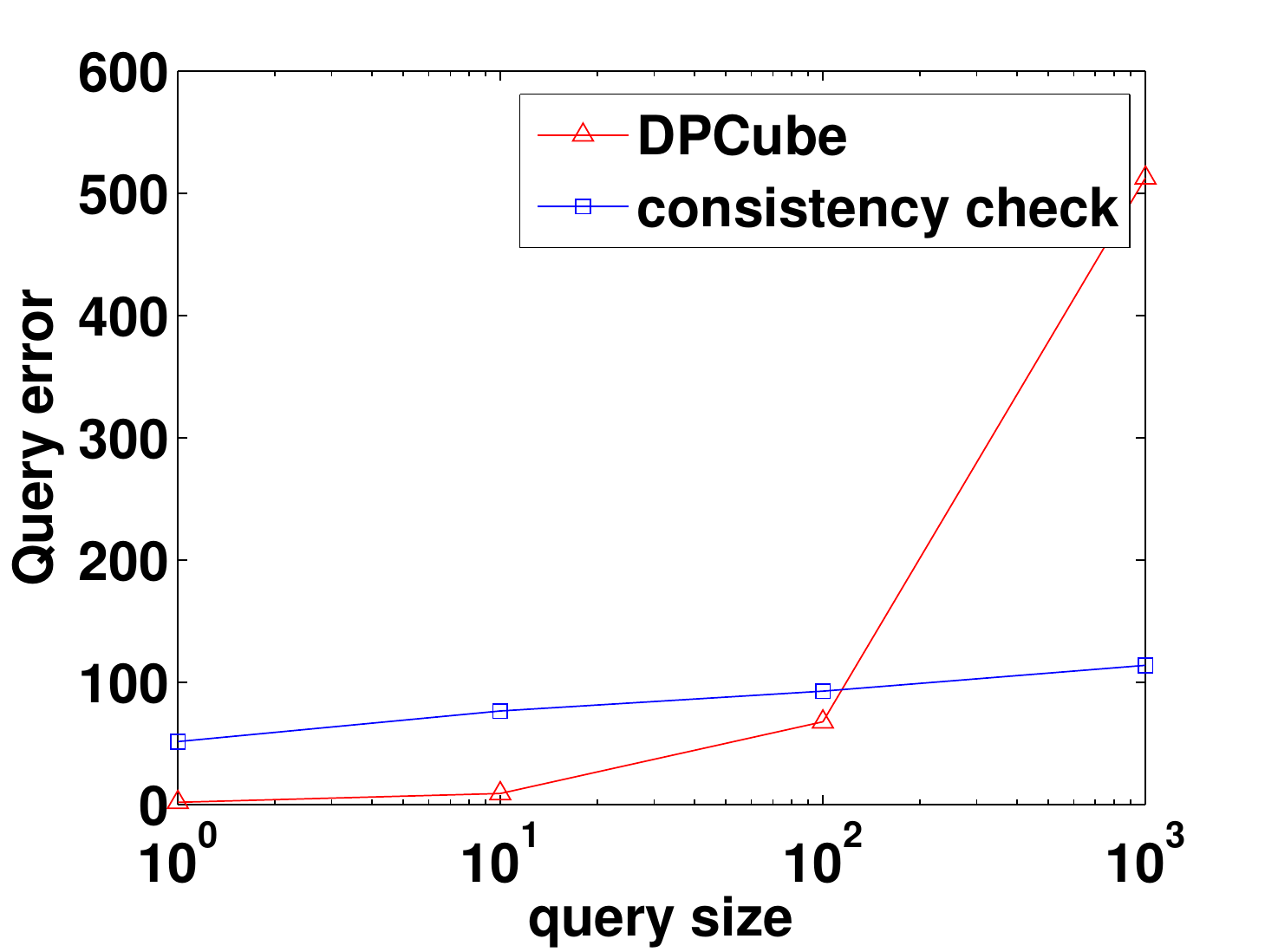}
\mbox{(b) vs. query size}
\end{minipage}
\caption{Query error for different approaches}
\label{Figure-error-comparison}
\end{figure}

\partitle{Comparison with Other Works}
We compare our approach with two representative approaches in existing works, the hierarchical kd-tree strategy used in \cite{inan10private}, and the hierarchical partitioning with consistency check in \cite{boost-accuracy}.
Figure \ref{Figure-error-comparison} shows the average absolute query error of different approaches with respect to varying privacy budget $\alpha$.  We can see that the DPCube algorithm achieves
best utility against the random query workload because of its efficient 2-phase use of the privacy budget and the v-optimal histogram.

We also experimented with query workloads with different query sizes for the DPCube approach and the consistency check approach to closely examine their differences.  Figure \ref{Figure-error-comparison}(b) shows the average absolute query error with respect to varying query sizes for the two approaches.  Note that the query size is in log scale.  As proven in our theoretical result in Figure \ref{Figure-E-error-LS-n}, the query error of our algorithm would increase with increasing query size.  We can see that for queries with reasonable sizes, our 2-phase algorithm achieves better results than the consistency check algorithm.  On the other hand, the consistency check algorithm is beneficial for large size queries because our algorithm favors smaller partitions with small variances which will result in large aggregated perturbation errors for large queries that spans multiple partitions.

\subsection{Additional Applications}
\label{sec-app}

\partitle{Classification}
We evaluate the utility of the released histogram for classification and compare it with other differentially private classification algorithms. In this experiment, the dataset is divided into training and test subsets with 30162 and 15060 records respectively.
We use the \emph{work class}, \emph{martial status}, \emph{race}, and \emph{sex} attributes as features.  The class was represented by \emph{salary} attribute with two possible values indicating if the salary is greater than \$50k or not.


For this experiment, we compare several classifiers.  As a baseline, we trained a \emph{ID3} classifier from the original data using Weka~\cite{Hall:2009}.
We also adapted the Weka ID3 implementation such that it can use histogram as its input.
To test the utility of the differentially private histogram generated from our algorithm, we used it to train an ID3 classifier called \emph{DPCube histogram ID3}.
As a comparison, we implemented an interactive differentially private ID3 classifier, \emph{private interactive ID3}, introduced by Friedman et al.~\cite{Friedman:2010}.

\begin{figure}[!htb]
  \begin{center}
  \includegraphics[width=4.5cm]{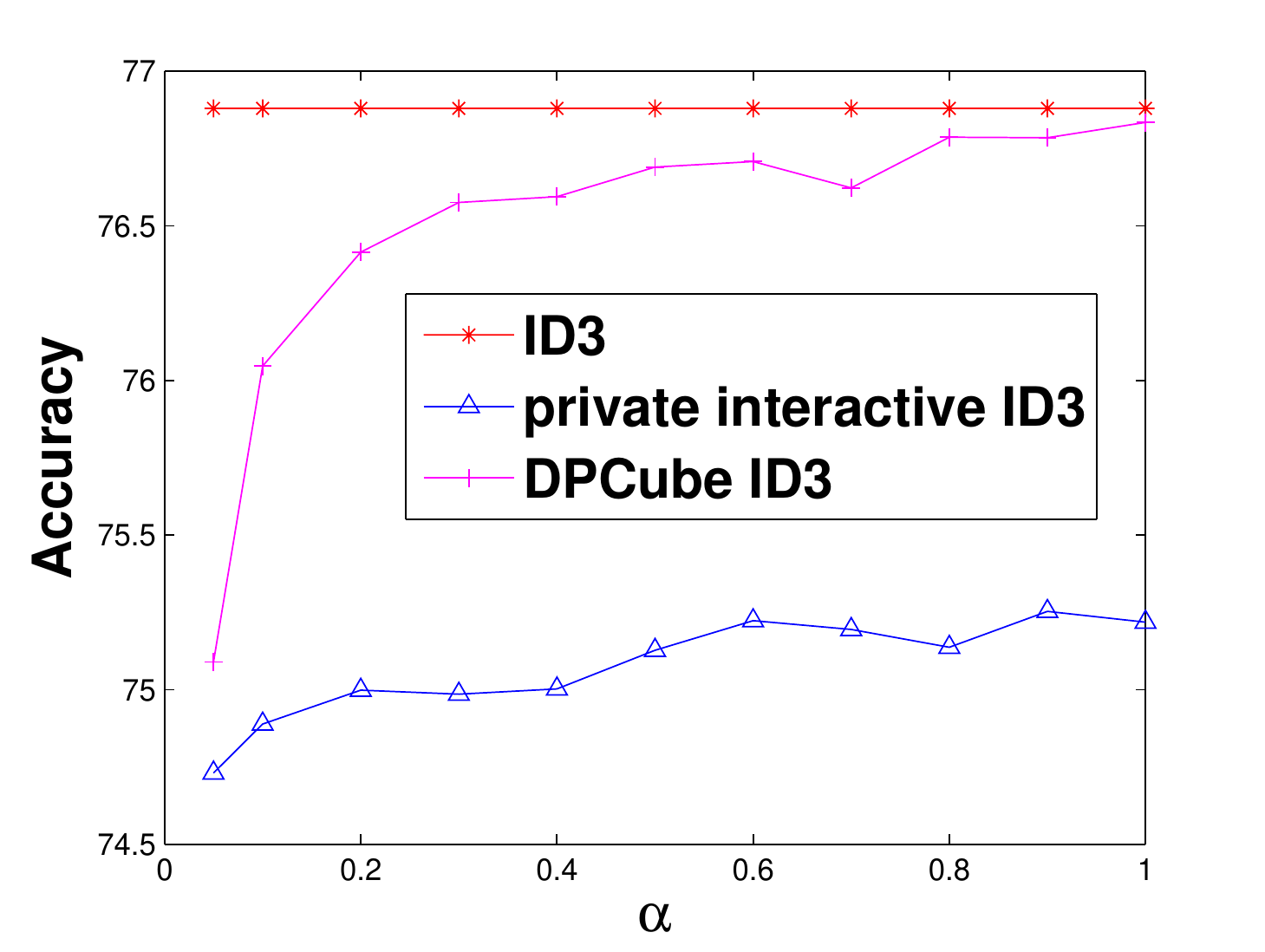}%
  \end{center}
  \vspace{-0.2cm}
  \caption{Classification accuracy vs. privacy budget $\alpha$}
  \label{fig:classifiers}
\end{figure}
Figure~\ref{fig:classifiers} shows the classification accuracy of the different classifiers with respect to varying privacy budget $\alpha$.
The original \emph{ID3} classifier provides a baseline accuracy at 76.9\%.
The \emph{DPCube ID3} achieves slightly worse but comparable accuracy than the baseline due to the noise. While both the DPCube ID3 and the private interactive ID3 achieve better accuracy with increasing privacy budget as expected, our DPCube ID3 outperforms the private interactive ID3 due to its efficient use of the privacy budget.

\partitle{Blocking for Record linkage}
We also evaluated the utility of the released histogram for record linkage and compared our method against the \emph{hierarchical kd-tree} scheme from \cite{inan10private}.
The attributes we considered for this experiment are: age, education, wage, marital status, race and sex.
As the histogram is used for the blocking step and all pairs of records in matching blocks will be further linked using an SMC protocol, our main goal is to reduce the total number of pairs of records in matching blocks in order to reduce the SMC cost.  We use the reduction ratio used in \cite{inan10private} as our evaluation metric.  It is defined as follows:

\begin{equation}
reduction\_ratio = 1 - \frac{\sum_{i=1}^k n_i * m_i}{n*m}
\end{equation}
where $n_i$ ($m_i$) corresponds to the number of records in dataset $1$ (resp. $2$) that fall into the $ith$ block, and $k$ is the total number of blocks.

\begin{figure}[h!]
\begin{minipage}{0.23\textwidth}
\centering
\includegraphics[width=4.5cm]{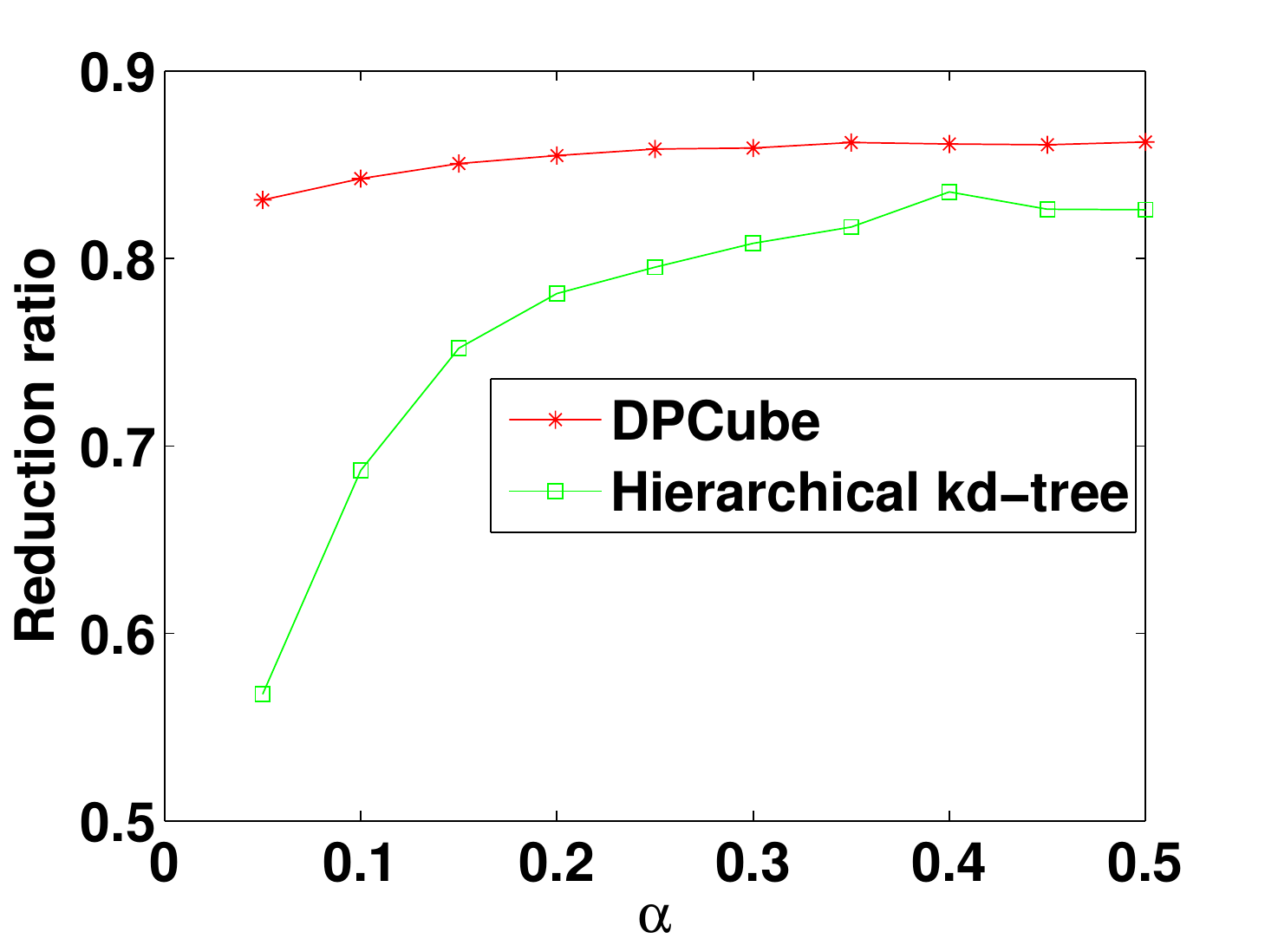}
\mbox{(a) vs. privacy budget $\alpha$}
\end{minipage}
\begin{minipage}{0.23\textwidth}
\centering
\includegraphics[width=4.5cm]{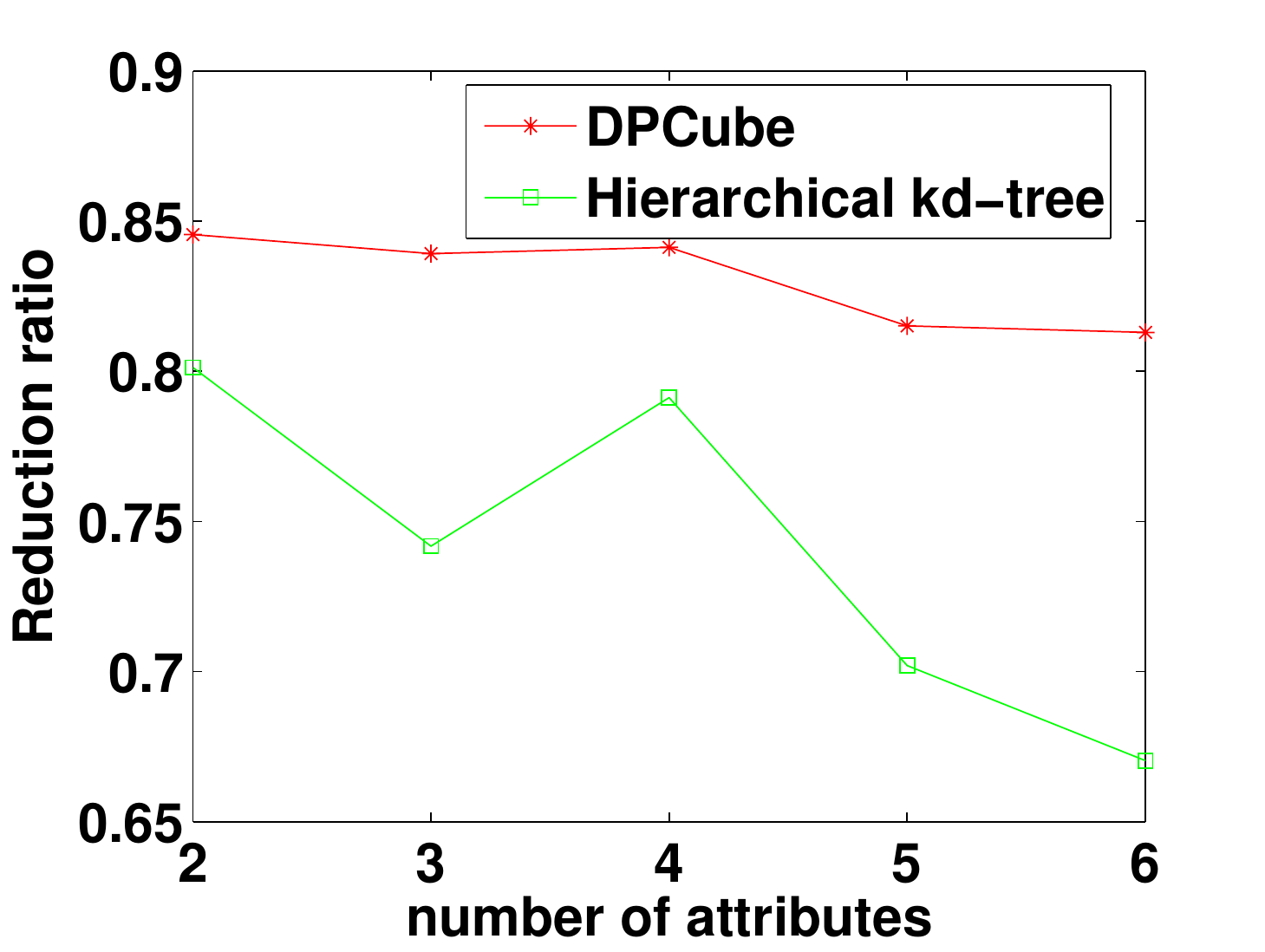}
\mbox{(b) vs. number of attributes}
\end{minipage}
\caption{Reduction ratio of blocking for record linkage}
\label{Figure-RL}
\end{figure}

We compared both methods by running experiments with varying privacy budget ($\alpha$) values (using the first 2 attributes of each record) and with varying numbers of attributes (with $\alpha$ fixed to 0.1).  Figure \ref{Figure-RL}(a) shows the reduction ratio with varying privacy budget.  Both methods exhibit an increasing trend in reduction ratio as the privacy budget grows but our 2-phase v-optimal histogram consistently outperforms the hierarchical kd-tree approach and maintains a steady reduction ratio around 85\%.
Figure \ref{Figure-RL}(b) shows the reduction ratio with varying number of attributes (dimensions).  As the number of attributes increases, both methods show a drop in the reduction ration due to the sparsification of data points, thus increasing the relative error for each cell/partition.   However, our DPCube approach exhibits desirable robustness when the dimensionality increases compared to the hierarchical kd-tree approach.

\section{Conclusions and Future works}

We have presented a two-phase multidimensional partitioning algorithm with estimation algorithms for differentially private histogram release.  
We formally analyzed the utility of the released histograms and quantified the errors for answering linear counting queries.  We showed that the released v-optimal histogram combined with a simple query estimation scheme achieves bounded query error and superior utility than existing approaches for ``smoothly" distributed data.  
The experimental results on using the released histogram for random linear counting queries and additional applications including classification and blocking for record linkage showed the benefit of our approach.
As future work, we plan to develop algorithms that are both data- and workload-aware to boost the accuracy for specific workloads and investigate the problem of releasing histograms for temporally changing data.

\bibliographystyle{IEEEtran}
\scriptsize{
\bibliography{ref/exportlist,ref/privacy}
}


\end{document}